\documentclass[12pt,a4paper]{article}
\pdfoutput=1
\usepackage{jheppub}
\usepackage{dsfont}
\usepackage{amssymb,amsmath}
\usepackage{epsfig}
\usepackage{epstopdf}
\usepackage{latexsym}
\usepackage{graphicx}
\usepackage{subfigure}
\usepackage{booktabs}
\usepackage{bbm}
\makeatletter
\def\@fpheader{\relax}
\makeatother

\def\O{{\cal O}}
\def\S{{\cal S}}
\def\A{{\cal A}}
\def\l{{\ell}}
\def\a{{\varepsilon}}

\def\T {{T_{\text{eff}}}}

\newcommand{\be}{\begin{equation}}
\newcommand{\ee}{\end{equation}}
\newcommand{\bea}{\begin{eqnarray}}
\newcommand{\eea}{\end{eqnarray}}

\subheader{\begin{flushright}
\end{flushright}}

\title{Aspects of Holographic Entanglement at Finite Temperature and Chemical Potential}
\author{Sandipan Kundu$^1$ and Juan F. Pedraza$^2$}
\affiliation{$^1$Department of Physics, Cornell University, Ithaca, NY 14853, USA}
\affiliation{$^2$Institute for Theoretical Physics, University of Amsterdam, 1090 GL Amsterdam, NL}
\emailAdd{kundu@cornell.edu}
\emailAdd{jpedraza@uva.nl}

\abstract{We investigate  the behavior of  entanglement entropy at finite temperature and chemical potential for  strongly coupled large-N gauge theories in $d$-dimensions ($d\ge 3$) that are dual to Anti-de Sitter-Reissner-Nordstrom geometries in $(d+1)-$dimensions, in the context of gauge-gravity duality. We develop systematic expansions based on the Ryu-Takayanagi prescription 
that enable us  to derive analytic expressions for entanglement entropy and mutual information in different regimes of interest. Consequently, we identify the specific regions of the bulk geometry that contribute most significantly to the entanglement entropy of the boundary theory at different limits. 
We define a scale, dubbed as the \emph{effective temperature}, which determines the behavior of entanglement in different regimes. At high effective temperature, entanglement entropy is dominated by the thermodynamic entropy, however, mutual information subtracts out this contribution
and measures the actual quantum entanglement. Finally, we study  the entanglement/disentanglement transition of mutual information in the presence of chemical potential which shows that the quantum entanglement between two sub-regions decreases with the increase of chemical potential.}

\begin{document}

\maketitle
\flushbottom

\section{Introduction}

Understanding the behavior of non-abelian gauge theories at  finite temperature and density is one of those classic problems that has intrigued physicists over the years. The problem is more challenging at strong coupling because the conventional perturbative field theory methods are markedly inadequate at strong coupling. In recent years, the strong-coupling physics has become even more relevant owing to the continued experimental explorations at the Relativistic Heavy Ion Collider (RHIC) and the Large Hadron Collider (LHC). Apart from possible applications, understanding the behavior of gauge theories at  finite temperature and chemical potential can provide useful insight into their dynamics. 

With the emergence of {\it holography} \cite{Banks:1996vh, Maldacena:1997re}, it has become clear that there is a remarkable connection between two of the cornerstones of theoretical physics: gauge theory and quantum gravity. The Anti-de-Sitter/Conformal Field Theory (AdS/CFT) correspondence \cite{Maldacena:1997re, Witten:1998qj, Gubser:1998bc, Aharony:1999ti} has been famously successful at providing a concrete realization of the idea of holography, leading to  theoretical control over a large class of strongly interacting quantum field theories. The AdS/CFT correspondence greatly simplifies  the computations of observables of certain strongly coupled large-N gauge theories in $d$-dimensions by translating them into classical gravity computations in $(d+1)$-dimensions.

Entanglement entropy is an important concept in quantum information theory which measures the quantum entanglement between two sub-systems of a given system.  Entanglement entropy has been used extensively in quantum field theories and quantum many body systems as a useful tool to characterize states of matter with long range correlations. For strongly coupled large-N gauge theories with holographic duals, there is an elegant proposal by Ryu and Takayanagi  for computing entanglement entropy: entanglement entropy associated to a region A is given by  the area of the bulk extremal surface anchored on the boundary of A \cite{Ryu:2006bv}. This simple yet powerful proposal  does satisfy several non-trivial relations \cite{Headrick:2007km,Headrick:2010zt} obeyed by entanglement entropy. In recent years, the Ryu-Takayanagi formula has been used extensively to analytically study entanglement entropy  in various holographic setups \cite{Fischler:2012ca,Fischler:2013fba,Caputa:2013lfa,Pang:2013lpa,Ben-Ami:2014gsa,Pang:2014tpa,Park:2015afa,Astaneh:2015gmg,Chaturvedi:2016kbk}.  In this article, our goal is to understand the behavior of  entanglement entropy at finite temperature and chemical potential for  strongly coupled large-N gauge theories in $d$-dimensions ($d\ge 3$) that are dual to AdS-Reissner-Nordstr\"{o}m geometries (AdS-RN) in $(d+1)-$dimensions. In order to achieve that we will develop systematic expansions using the Ryu-Takayanagi formula, leading to analytic expressions for entanglement entropy in different regimes.   Beyond possible applications, the study of entanglement entropy at finite temperature and chemical potential using the AdS/CFT correspondence is an interesting problem on its own right and it can provide new insight into the nature of quantum entanglement at strong coupling.

The AdS/CFT   correspondence bestows another side to the problem which is even more intriguing. The AdS/CFT correspondence has tempted us with its potentiality of addressing the questions of quantum gravity by translating them into gauge theory language. In recent years, a great deal of attention has been focused on understanding how the gravitational degrees of freedom emerges from quantum entanglement \cite{Lashkari:2013koa, Faulkner:2013ica, Swingle:2014uza}. For that it is crucial to identify the specific locations of the `bulk' gravity theory that contribute most significantly to the entanglement entropy of the `boundary' field theory at different limits. For example, at high temperature (and/or chemical potential), we will show that the most dominant contribution to the entanglement entropy comes from the near horizon part of the bulk geometry and hence, to learn about the physics in the vicinity of the horizon, an analytic understanding of the holographic entanglement entropy will be very useful.

\begin{figure}[htbp]
\begin{center}
\includegraphics[height=4.1cm]{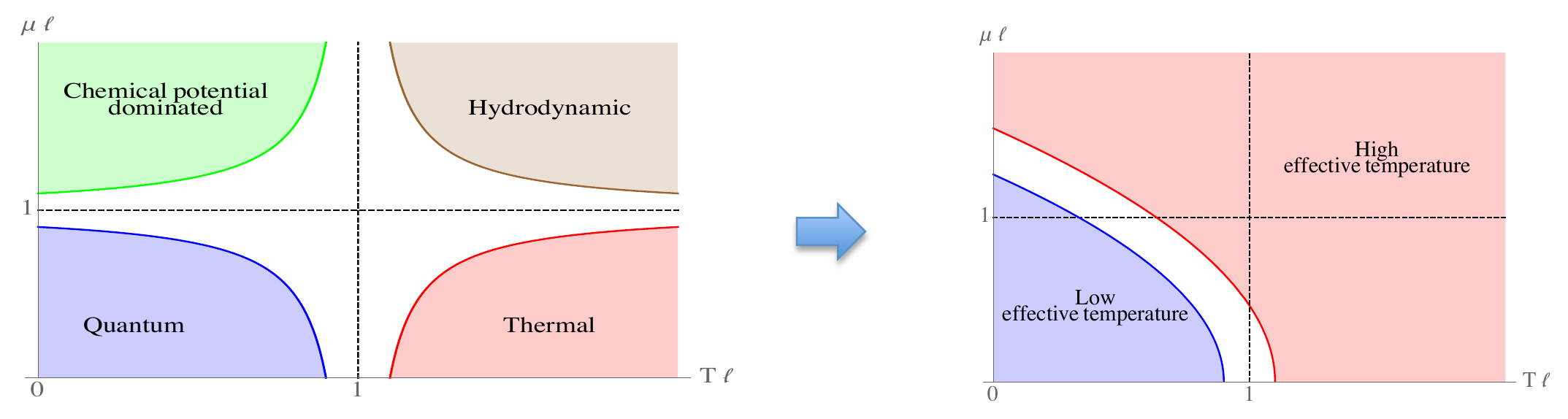}
\caption{For quantum field theories at finite temperature and chemical potential, the behavior of entanglement entropy associated with a region of size $\l$ is controlled by different physics in different regimes, as shown in the left hand side figure. We show that for theories with holographic dual descriptions, effective temperature $\T$ completely determines the behavior of entanglement entropy and hence  it is sufficient to consider (i) high effective temperature regime and (ii) low effective temperature regime, as shown in the right hand side figure.}
\label{regimes4}
\end{center}
\end{figure}

For quantum field theories at finite temperature and chemical potential, entanglement entropy associated with a region of size $\l$ can be used to explore different regimes that are controlled by different physics: (a) quantum ($\mu\l<<1, T\l<<1$), (b) thermal ($\mu\l<<1, T\l>>1$), (c) chemical potential dominated ($\mu\l>>1, T\l<<1$) and (d) hydrodynamic ($\mu\l>>1, T\l>>1$). For holographic theories, we will extend the analytic techniques developed in \cite{Fischler:2012ca} to explore the behavior of entanglement entropy in all these regimes.\footnote{Recently, holographic entanglement entropy at finite temperature and chemical potential for $(2+1)-$dimensional boundary theory has been studied in \cite{Chaturvedi:2016kbk}.} We will show that for holographic theories it is more convenient to label the state of the field theory in terms of an effective temperature $\T(\mu,T)$ and a dimensionless energy parameter $\varepsilon(\mu,T)$. The effective temperature $\T$ is defined such that the entropy density of the system goes as $s\sim \T^{d-1}$ and hence $\T$ counts the number of microstates of the system for a particular temperature and chemical potential.\footnote{$\T$ interpolates between $\T\propto T$ and $\T\propto\mu$ as one goes from $\mu/T\ll1$ to $\mu/T\gg1$. We will discuss this in more details in the next section.} Whereas,  the energy density is proportional to the energy parameter $\varepsilon\sim \O(1)$ for all the macrostates of the system with the same number of microstates. A reasonable expectation is that $\T$ will play a crucial role in determining the behavior of entanglement entropy in different regimes (see figure \ref{regimes4}). We will confirm this guess by performing some explicit calculations. In particular, at low effective temperature, i.e. $\T\ll 1/\l$, the extremal surface is restricted to be near the boundary region and hence the most dominant contribution to the entanglement entropy comes from the AdS-boundary. This leading contribution is just the entanglement due to vacuum quantum fluctuations. The corrections terms are due to the deviation of the bulk geometry from pure AdS. At low effective temperature, these corrections are small and can be computed perturbatively. On the other hand, at high effective temperature, i.e. $\T\gg 1/l$, the contributions of finite temperature and/or chemical potential to the entanglement entropy become more and more significant and in the dual gravity theory the extremal surface associated with the entangling region approaches the horizon exponentially fast but always stays at a finite distance above the horizon \cite{Hubeny:2012ry,Fischler:2012ca}.\footnote{At $T=0$ but finite $\mu$, the AdS-RN becomes extremal and in this case the extremal surface approaches the horizon only at a power law rate in the limit $\mu\l\gg1$ (see appendix \ref{approach}).} At high effective temperature, the extremal surface tends to wrap a part of the horizon and the leading finite contribution to the entanglement entropy comes from the near horizon region of the bulk and it is just the thermodynamic entropy. Whereas the full bulk geometry contributes to the sub-leading terms which actually measure quantum entanglement between the region and its surroundings and hence these sub-leading terms contain far more interesting information about these gauge theories.

In quantum information theory, an important quantity derived from entanglement entropy is mutual information.  Mutual information between two disjoint sub-systems A and B is defined as $I(A,B)=S_A+S_B- S_{A\cup B}$, where $S_A$, $S_B$ and $S_{A\cup B}$ denote entanglement entropy of the region $A$, $B$ and $A\cup B$ respectively with the rest of the system. Entanglement entropy of a spatial region in a local field theory is UV-divergent. Only local physics contributes to the UV-divergent piece, whereas the finite part contains information about the long range entanglement. Mutual information is a more useful quantity because it has several advantages over entanglement entropy. It is (i) UV-finite, (ii) positive semi-definite and (iii) a measure of the total correlation between the two sub-systems: including both classical and quantum correlations \cite{PhysRevLett.100.070502}. We will use our analytic results of entanglement entropy to compute mutual information between two disjoint regions in different regimes of interest. At high effective temperature, entanglement entropy is dominated by the thermodynamic entropy \cite{Tonni:2010pv}.
It was pointed out in \cite{Fischler:2012uv}, that at finite temperature mutual information subtracts out the thermal part of the entanglement entropy and satisfies an area law.
We will show that the same is true even in the presence of chemical potential and hence mutual information indeed measures the actual quantum entanglement between two sub-regions. It is also known that mutual information undergoes an entanglement/disentanglement ``phase-transition" for large N gauge theories which have holographic dual descriptions \cite{Headrick:2010zt,Tonni:2010pv,Fischler:2012uv,MolinaVilaplana:2011xt,Fischler:2013gsa}.
We will show that at finite chemical potential mutual information undergoes an entanglement/disentanglement
transition reminiscent to the phase transition at finite temperature but zero chemical potential. Our results clearly demonstrates that the quantum entanglement between two sub-regions decreases with increasing effective temperature.
This implies that, at a fixed temperature, entanglement decreases with increasing chemical potential.

The rest of the paper is organized as follows. In section \ref{sectemp} we start with a brief review of holographic gauge theories at finite temperature and chemical potential and introduce effective temperature. In section \ref{sectionee}, we develop a systematic expansion for entanglement entropy at high and low effective temperature. Then using that expansion we examine the behavior of entanglement entropy in different regimes in section \ref{section4}.  In section \ref{sectionmi}, we use the results of the previous section to obtain analytic expressions for mutual information. In this section, we also study the entanglement/disentanglement phase transition of mutual information in the presence of chemical potential. Then we conclude in section \ref{conclusions} with future directions.

\section{Gauge theories at finite temperature and chemical potential}\label{sectemp}

In this paper, we will consider strongly coupled large-$N$ gauge theories in $d-$dimensions that are dual to AdS$_{d+1}$ with $d\ge 3$. Furthermore, we also want an additional  global U(1) symmetry which is very common in condensed matter systems. The AdS/CFT correspondence has taught us that a global U(1) symmetry can be introduced in the boundary field theory by adding a Maxwell field to the bulk spacetime. The physics at finite temperature and chemical potential (corresponding to the global U(1)) is described by the  AdS-RN$_{d+1}$ black hole solutions.

Consider the $(d+1)-$dimensional Einstein-Hilbert action wth a negative cosmological constant coupled to a Maxwell field,
\begin{eqnarray} \label{action1}
S = \frac{1}{8\pi G_N^{(d+1)}}\left(\frac{1}{2} \int d^{d+1} x \sqrt{-g} \left(R - 2 \Lambda \right) - \frac{1}{4} \int d^{d+1}x \sqrt{-g} F_{\mu\nu} F^{\mu\nu}
\right) \ ,
\end{eqnarray}
with $\Lambda = -\frac{d(d-1)}{2 L^2}$. The above action gives the following equations of motion
\begin{eqnarray}
&& R_{\mu\nu} - \frac{1}{2} \left(R- 2 \Lambda \right) g_{\mu\nu} = g^{\alpha\rho} F_{\rho\mu} F_{\alpha\nu} - \frac{1}{4} g_{\mu\nu} \left(F^{\alpha\beta}
F_{\alpha \beta}\right) \ , \label{eom1} \\
&& \partial_\rho \left[ \sqrt{-g} g^{\mu\rho} g^{\nu\sigma} F_{\mu\nu} \right] = 0 \ . \label{eom2}
\end{eqnarray}

\subsection{Equilibrium solutions: AdS-RN}

There is a family of two-parameter black hole solutions of (\ref{eom1})-(\ref{eom2}) known as the AdS-Reissner-Nordstr\"{o}m black holes \cite{Cvetic:1999ne,Chamblin:1999tk}. For $d\ge 3$ the solutions are:
\begin{eqnarray} \label{RN}
&& ds^2 = \frac{L^2}{z^2} \left(- f(z) dt^2 + \frac{dz^2}{f(z)} + d\vec{x}^2 \right) \ , \nonumber\\
&& f(z) = 1- M z^d + \frac{(d-2)Q^2}{(d-1)L^2 } z^{2(d-1)} \ ,\\
&& A_t = Q (z_H^{d-2} - z^{d-2}) \nonumber\ ,
\end{eqnarray}
where $M$ is the mass of the black hole and $Q$ is the charge. Here, $z_H$ denotes the location of the horizon which is given by the smallest real root of
$f(z)=0$.
The dual theory is a CFT that lives in $d$ spacetime dimensions and is characterized by a  density matrix in the grand canonical ensemble, $\rho = e^{-\beta (H -\mu q)}$, where $q$ is the total charge. From the point of view of the boundary, having a non-zero electric field in the bulk corresponds to
having a chemical potential which, for $d\geq3$, is given by
\begin{eqnarray}
\mu \equiv \frac{1}{L}\lim_{z\to 0} A_t(z) = \frac{Q}{L} z_H^{d-2} \ .
\end{eqnarray}

In this paper, we will only consider field theories that live in spacetime dimensions $d\ge 3$. The temperature of the dual field theory can be identified as the Hawking temperature of the black hole. For  $d\geq3$ we have
\be
T=-\frac{1}{4\pi}\frac{d}{dz}f(z)\bigg|_{z_H}=\frac{d}{4 \pi z_H}\left(1-\frac{(d-2)^2 Q^2 z_H^{2(d-1)}}{d(d-1)L^2}\right) \ .
\ee
It is clear that, for $Q^2=d(d-1)L^2/(d-2)^2z_H^{2(d-1)}$, the black hole is extremal and the dual field theory is then at zero temperature but with finite
chemical potential. This solution can be written as follows:
\be
f(z) = 1- \frac{2(d-1)}{d-2}\left(\frac{z}{z_H}\right)^d + \frac{d}{d-2} \left(\frac{z}{z_H}\right)^{2(d-1)} \ ,
\ee
with
\be
\mu=\frac{1}{z_H}\sqrt{\frac{d (d-1) }{(d-2)^2}} \ .
\ee

\subsection{Effective temperature}
It is useful to write down the metric (\ref{RN}) in the following form
\be\label{fofznew}
f(z) = 1- \varepsilon \left(\frac{z}{z_H}\right)^d + \left(\varepsilon-1\right) \left(\frac{z}{z_H}\right)^{2(d-1)}\ .
\ee
Chemical potential and temperature are now given by
\be\label{muandT}
\mu=\frac{1}{z_H}\sqrt{\frac{ (d-1)}{(d-2)}(\varepsilon-1)} \ , \qquad T=\frac{ 2(d-1)-(d-2)\varepsilon}{4\pi z_H}\ .
\ee
In this parametrization, it can be easily shown that $1\ge\varepsilon\ge \frac{2(d-1)}{d-2}$ and
\be\label{delta1}
\varepsilon(T,\mu)=a-\frac{2b}{1+\sqrt{1+\frac{d^2}{2\pi^2 a b}\left(\frac{\mu^2}{T^2}\right)}}\ ,
\ee
where $a$ and $b$ depend only on spacetime dimensions
\be\label{abdefs}
a=\frac{2(d-1)}{d-2}\ , \qquad b=\frac{d}{d-2}\ .
\ee
Note that $a-b=1$. We will also define an effective temperature $\T(T,\mu)$, which will play a crucial role
\begin{align}\label{Teff1}
\T(T,\mu) \equiv \frac{d}{4\pi z_H}=\frac{T}{2}\left[1+\sqrt{1+\frac{d^2}{2\pi^2 a b}\left(\frac{\mu^2}{T^2}\right)}\right]\ .
\end{align}
Before we proceed, let us make some comments on the physical importance of these two parameters: $\T$ and $\varepsilon$. We will show later that the entropy density of the system goes as $s\sim \T^{d-1}$ and hence $\T$ counts the number of microstates of the system for a particular temperature and chemical potential. It is also very reasonable to expect that $\T$ will play a crucial role in determining the behavior of entanglement entropy in different regimes. On the other hand, $\varepsilon\sim \O(1)$ is a dimensionless quantity which measures the energy of the system. More precisely, we will show that for all the macrostates of the system with the same number of microstates, the energy density is proportional to $\varepsilon$. Also note that both $\T$ and $\varepsilon$ are monotonically increasing functions of $\mu/T$. Let us now note some special cases:
\begin{itemize}
\item {Zero temperature and chemical potential: this is the special case $z_H\rightarrow \infty$.}
\item {Finite temperature and zero chemical potential: this corresponds to the case $\varepsilon=1$ and
\be
T=\T=\frac{d}{4\pi z_H}\ .
\ee
 }
\item {Zero temperature and finite chemical potential: this corresponds to the case $\varepsilon=a$ with
\be
\mu=\frac{\sqrt{d(d-1)}}{(d-2)z_H}\ , \qquad \T=\frac{\mu d (d-2)}{2\pi\sqrt{2d(d-1)}}\ .
\ee
}
\end{itemize}
More generally, from (\ref{Teff1}) it follows that $\T$ interpolates between $\T\propto T$ and $\T\propto\mu$ as one goes from $\mu/T\ll1$ to $\mu/T\gg1$. Specifically, for $\mu/T\ll1$ we have
\be\label{zhthermal}
\T=T\left[1+\frac{d^2}{8\pi^2ab}\left(\frac{\mu^2}{T^2}\right)+\mathcal{O}\left(\frac{\mu^4}{T^4}\right)\right]\,.
\ee
In the opposite limit $\mu/T\gg1$ we find
\be\label{zhquantum}
\T=\frac{d-2}{4\pi }\sqrt{\frac{2b}{a}}\mu\left[1+\frac{2\pi }{d-2} \sqrt{\frac{a}{2b}}\left(\frac{T}{\mu
}\right)+\mathcal{O}\left(\frac{T^2}{\mu^2}\right)\right]\,.
\ee

\subsection{Stress-energy tensor}

Any asymptotically AdS metric can be written in the Fefferman-Graham form \cite{fg}
\begin{equation}\label{feffermangraham}
ds^2={L^2 \over z^2}\left(g_{\mu\nu}(z,x)dx^{\mu}dx^{\nu}+dz^2\right)~.
\end{equation}
The function $g_{\mu\nu}(z,x)$ encodes data dual to the CFT metric $ds^2=g_{\mu\nu}(0,x) dx^{\mu}dx^{\nu}$ and the expectation value of the CFT stress-energy tensor $T_{\mu\nu}(x)$.
More specifically, in terms of the near-boundary expansion
\begin{equation}\label{metricexpansion}
g_{\mu\nu}(z,x)=g_{\mu\nu}(x)+z^2 g^{(2)}_{\mu\nu}(x)+\ldots +z^d g^{(d)}_{\mu\nu}(x)+ z^d \log (z^2) h^{(d)}_{\mu\nu}(x)+\ldots~,
\end{equation}
the standard GKPW prescription for correlation functions \cite{Gubser:1998bc,Witten:1998qj} after appropriate holographic renormalization leads to
\cite{dhss,skenderis,skenderis2}
\begin{equation}\label{graltmunu}
\left\langle T_{\mu\nu}(x)\right\rangle={ d\, L^{d-1} \over 16\pi G^{(d+1)}_{N}}\left(g^{(d)}_{\mu\nu}(x)+X^{(d)}_{\mu\nu}(x)\right)~,
\end{equation}
where $X^{(d)}_{\mu\nu}=0$ $\forall$ odd $d$ and\footnote{This is a reflection of the fact that there are no gravitational conformal anomalies in odd boundary dimensions.}  
\begin{eqnarray}\label{xmunu}
X^{(2)}_{\mu\nu}&=&-g_{\mu\nu}g^{(2)\alpha}_{\alpha}~,\\
X^{(4)}_{\mu\nu}&=&-{1\over 8}g_{\mu\nu}\left[\left(g_{\alpha}^{(2)\alpha}\right)^2
-g_{\alpha}^{(2)\beta}g_{\beta}^{(2)\alpha}\right]
-{1\over 2}g_{\mu}^{(2)\alpha}g_{\alpha\nu}^{(2)}
+{1\over 4}g^{(2)}_{\mu\nu}g_{\alpha}^{(2)\alpha}~,\nonumber
\end{eqnarray}
and $X^{(2d)}_{\mu\nu}$ for $d\geq3$ given by similar expressions that we will not transcribe here. In (\ref{xmunu}) the indices
of the tensors $g^{(n)}_{\mu\nu}(x)$ are raised with the inverse boundary metric $g^{\mu\nu}(x)$.

In order to obtain the stress-energy tensor from the AdS-RN$_{d+1}$ metric (\ref{RN}) we have to write it in the Fefferman-Graham form (\ref{feffermangraham}).
This can be done perturbatively in a near-boundary expansion. Specifically, after the coordinate transformation
\be
z=\tilde{z}\left(1-\frac{\varepsilon \tilde{z}^d}{2 d z_H^d}+\O(\tilde{z}^{2(d-1)})\right)\,,
\ee
we arrive at
\begin{equation}
ds^2={L^2 \over \tilde{z}^2}\left[(\eta_{\mu\nu}+\tau_{\mu\nu}\tilde{z}^d+\O(\tilde{z}^{2{d-1}}))dx^\mu dx^{\nu}+d\tilde{z}^2\right]~,
\end{equation}
where
\bea
\tau_{00}=\frac{(d-1)\varepsilon}{d z_H^d}\,,\qquad \tau_{ii}=\frac{\varepsilon}{d z_H^d}\,.
\eea
From here it follows that the stress-energy tensor for $d$-dimensional boundary theory dual to AdS-RN$_{d+1}$ is given by:
\bea\label{epsilondef}
\langle T_{00}\rangle &\equiv&\mathcal{E}=\frac{L^{d-1} (d-1)}{16\pi G^{(d+1)}_{N}}\left(\frac{4 \pi  \T}{d}\right)^d\varepsilon\ ,\\
\langle T_{ii}\rangle &\equiv&P=\frac{L^{d-1}}{16\pi G^{(d+1)}_{N}}\left(\frac{4 \pi  \T}{d}\right)^d\varepsilon\ .
\eea
In particular, notice that the energy density and pressure satisfy $\mathcal{E} = (d-1)P$ making the stress tensor traceless, as expected for a CFT. As mentioned earlier, for all the macrostates of the system with the same $\T$, i.e., the same number of microstates, the energy density is proportional to $\varepsilon$. Also note that since both $\T$ and $\varepsilon$ are monotonically increasing functions of $\mu/T$, energy density and pressure of the system at a fixed $T$ increase with increasing chemical potential.

\subsection{Thermodynamics}

We can also compute the various other thermodynamic quantities in terms of $T$ and $\T$. The entropy density of the dual CFT can be computed from the
Bekenstein-Hawking formula for black hole entropy $S_{\text{BH}}=A/4G^{(d+1)}_{N}$, where $A$ is the area of the horizon. The horizon lies at $z=z_H$ and $t=$ constant slice, and has an `area' of
\be
A=\int d^{d-1}x\sqrt{g}=\frac{L^{d-1}}{z_H^{d-1}}\text{Vol}(\mathbb{R}^{d-1})\,,
\ee
where $\text{Vol}(\mathbb{R}^{d-1})=\int d^{d-1}x$ is the (infinite) volume spanned in the $\vec{x}$-directions. Thus, the entropy density is given by
\be
s=\frac{L^{d-1}}{4G^{(d+1)}_{N}}\left(\frac{4 \pi  \T}{d}\right)^{d-1}\,.
\ee
Therefore, $\T$ indeed counts the number of microstates of the system for a particular temperature and chemical potential. Also note that the entropy for these holographic systems is non-zero even for $T=0$, which suggests that the ground state is degenerate.\footnote{One can argue that this state shouldn't be the true ground state. For instance, there are electron star solutions that are actually favored (less free energy) than the extremal black holes. This issue of finite entropy at $T=0$ has also been discussed in \cite{Horowitz:1996qd,Carroll:2009maa}.}

Similarly, the chemical potential and charge density are given by:
\be
\mu=\sqrt{\frac{(d-1)}{(d-2)}(\varepsilon-1)}\left(\frac{4 \pi  \T}{d}\right)\ ,
\ee
and
\be
\rho=\frac{L^{d-1}(d-2)}{8 \pi  G^{(d+1)}_{N}}\sqrt{\frac{(d-1)}{(d-2)}(\varepsilon-1)}\left(\frac{4 \pi  \T}{d}\right)^{d-1}\,,
\ee
respectively. Together, they satisfy the first law of thermodynamics:
\be
d\mathcal{E}=Tds+\mu d\rho\,.
\ee
Also note that charge density can be written in terms of only chemical potential and the effective temperature: $\rho\sim\mu \T^{d-2}$.

\section{Holographic entanglement entropy}\label{sectionee}
\begin{figure}[htbp]
\begin{center}
 \fbox{\includegraphics[height=5cm]{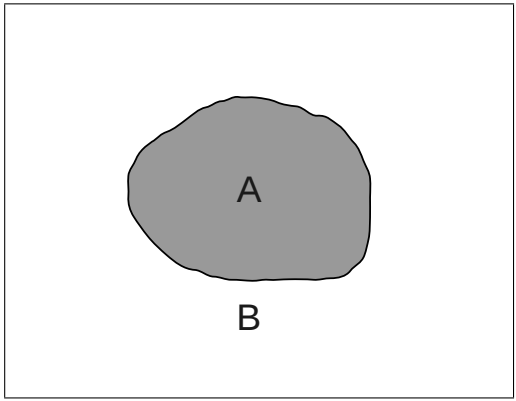}}
\caption{The total system is divided into two subsystems $A$ and $B$; the entanglement entropy measures the amount of information loss because of tracing over
the degrees of freedom of $B$.}
\label{ss}
\end{center}
\end{figure}
We are interested in computing entanglement entropy in the boundary field theory. The state of the system is completely specified by its density matrix $\rho$, a self-adjoint, positive semi-definite, trace class operator. Let us now imagine an entangling  surface  that  divides  the  entire  system in two subsystems, $A$ and its complement $B$, (see figure \ref{ss}) so that the total Hilbert space factorizes as $\mathcal{H}_{\text{total}}=\mathcal{H}_A\otimes\mathcal{H}_B$. The entanglement entropy of the region $A$ is defined as the von Neumann entropy,
\begin{equation}
S_A = - {\rm tr}_A \, \rho_A \log \rho_A \ .
\end{equation}
Here, $\rho_A$ is the reduced density matrix, obtained by tracing over the degrees of freedom of $B$:  $\rho_A = {\rm tr}_{B} \, \rho$.  Entanglement entropy is a highly non-local quantity and hence it could in principle measure quantum correlations which are not accessible to other observables constructed from any subset of local operators.

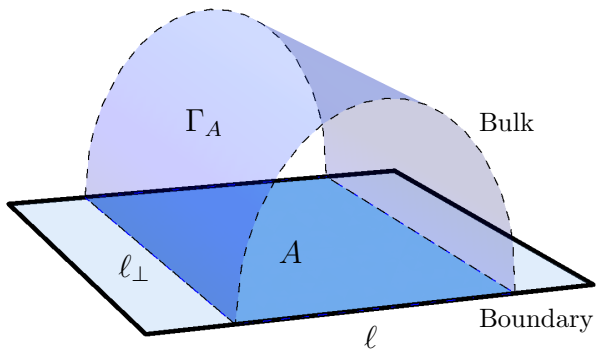
\begin{figure}[htbp]
\begin{center}
  \includegraphics[angle=0,width=0.5\textwidth]{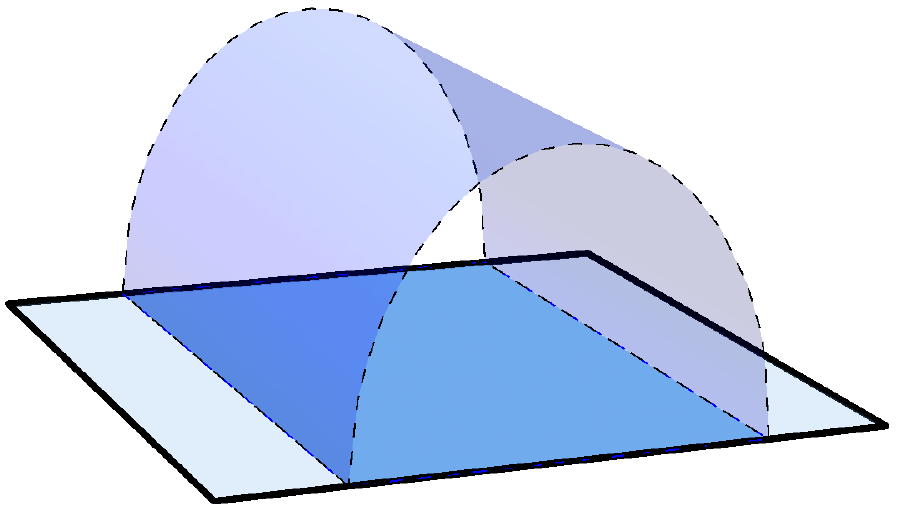}
\begin{picture}(0,0)
\put(-125,30){$A$}
\put(-93,-2){$\ell$}
\put(-185,25){$\ell_\perp$}
\put(-160,80){$\Gamma_A$}
\put(-50,6){\footnotesize{Boundary}}
\put(-50,80){\footnotesize{Bulk}}
\end{picture}
\end{center}
\vspace{-0.5cm}
\caption{\small A schematic diagram of the holographic prescription of the entanglement entropy.}
\label{figthinshell}
\end{figure}

In the context of the AdS/CFT correspondence, Ryu and Takayanagi \cite{Ryu:2006bv} proposed the following elegant prescription to compute entanglement entropy of a region $A$:
\be \label{rt}
S_A = \frac{1}{4 G_N^{(d+1)}} {\rm min} \left[ {\rm Area} \left(\Gamma_A \right)\right] \ ,
\ee
where $G_N$ is the bulk Newton's constant, and $\Gamma_A$ is a $(d-1)$-dimensional surface such that $\partial \Gamma_A = \partial A$ (see figure \ref{figthinshell} for a schematic representation). 

We will compute the entanglement entropy in the boundary theory for an infinite rectangular strip specified by
\begin{equation}
x\equiv x^1 \in \left[-\frac{\ell}{2},\frac{\ell}{2}\right],~  x^i\in \left[-\frac{\ell_\perp}{2},\frac{\ell_\perp}{2}\right],\qquad i=2,...,d-2
\end{equation}
with $\ell_\perp \rightarrow \infty$. The extremal surface $\Gamma_A$ is invariant under translations in $x^i, i=2,...,d-2$ and hence without loss of generality,
we can parameterize it with a single function $z(x)$ and boundary conditions
\be\label{bc}
z(\pm\ell/2)=\epsilon\ ,
\ee
where $\epsilon$ is a radial UV cutoff.

\subsection{Regimes of interest}
Before proceeding further, let us elaborate on the physical interpretation of the regimes that we will consider. In general, we have two independent parameters
that determine the state of the CFT, the temperature $T$ and the chemical potential $\mu$. In addition, we have another variable that we can tune: the size of
the entangling region $\ell$.

\begin{figure}[htbp]
\begin{center}
\includegraphics[height=7cm]{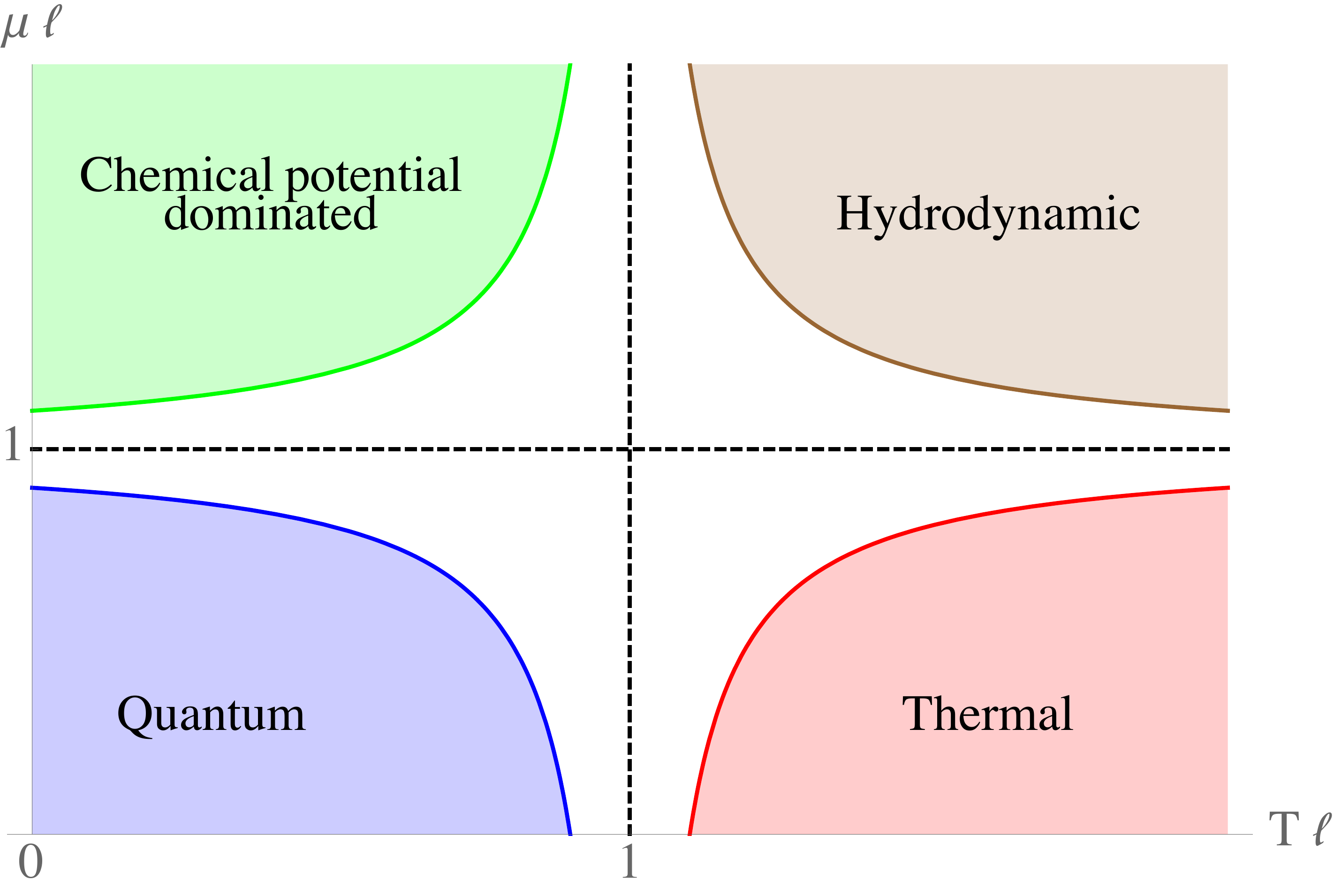}
\caption{For quantum field theories at finite temperature and chemical potential, the behavior of entanglement entropy associated with a region of size $\l$ is controlled by different physics in different regimes: (i) quantum, (ii) thermal, (iii) chemical potential dominated and (iv) hydrodynamic.}
\label{regimes1}
\end{center}
\end{figure}

From these three parameters $\{T,\mu,\ell\}$ it is possible to construct two dimensionless quantities, for which we have a number of possibilities. The first parameter we will consider is $T\l$.
This variable is rather intuitive, since it measures the strength of thermal fluctuations. Two regimes $T\l \gg1$ and $T\l\ll1$ are distinct and the behavior of entanglement entropy depends on the nature of the the relevant excitations, thermal or quantum mechanical, respectively.

For the second parameter we will use $\mu\l$. If $\mu\l\ll1$, the subregion we are focusing on is too small to be affected by the presence of chemical potential. In that case, the dominant contribution to entanglement entropy comes from vacuum fluctuations for $T\l\ll 1$ and from thermal fluctuations for $T\l\gg 1$. For, $\mu \l \gg1$, chemical potential plays a significant role since the subsystem is large in comparison to $1/\mu$. If $\mu \l \gg1$ and $T\l\ll1$, the physics is entirely controlled by chemical potential. Whereas in the limit $\mu \l \gg1$ and $T\l\gg1$, the system exhibits a hydrodynamic regime where entanglement entropy is dominated by the thermodynamic entropy. Thus, in total we have four possible regimes (see figre \ref{regimes1}), which are summarized in the Table 1 below.
\begin{table}[h!]
\begin{center}
\begin{tabular}{|c|c|c|}
 \hline
$\mu \l$ $\backslash$ $T\l$ & $T\l\ll1$  & $T\l\gg1$  \\
\hline
 $\mu\l\ll1$ & Quantum & Thermal  \\
\hline
 $\mu\l\gg1$ & Chemical potential dominated & Hydrodynamic  \\
\hline
\end{tabular}
\caption{Regimes of interest for the entanglement entropy computation.}
\label{table1}
\end{center}
\end{table}

For holographic theories, $\T$ is the most relevant scale of the theory  and $\T$ completely determines the behavior of entanglement entropy. Hence, it is sufficient to only consider (i) high effective temperature regime and (ii) low effective temperature regime, a fact that will be very useful below. It is known that for holographic theories even the $\mu/T\to\infty$ limit exhibits a hydrodynamic regime when the frequency and momentum of the excitations are small in comparison to the chemical potential  (see for instance \cite{Edalati:2009bi,Edalati:2010hk,Edalati:2010pn}). We will show that the same is true when $\mu\l\gg1$ even at zero temperature. This is a consequence of the fact that $\mu\l\gg1$ corresponds to $\T\l\gg 1$.

\subsection{Entanglement entropy expansion for $d\ge 3$}

Having established the regimes of interest, let us now write down the area functional for the rectangular infinite strip\footnote{We will set the AdS radius $L=1$. We will restore $L$ in the final results by dimensional analysis.}
\be\label{larangianStatic}
\mathcal{A}=2\ell_\perp^{d-2}\int_{0}^{\ell/2} \frac{dx}{z^{d-1}} \sqrt{1 + \frac{z'^2}{f(z)}}  \ ,
\ee
and the corresponding equation of motion reduces to
\be\label{coneq2}
1 + \frac{z'^2}{f(z)} = \left(\frac{z_*}{z}\right)^{2(d-1)} \ ,
\ee
where, $z_*$ is an integral of motion and $z=z_*$ represents the point of closest approach of the extremal surface. Each extremal surface has two branches,
joined smoothly at $(z=z_*, x=0)$ and $z_*$ can be determined from the boundary conditions (\ref{bc}). Now plugging (\ref{coneq2}) back into
(\ref{larangianStatic}) we get the area of the extremal surface
\be\label{onshellA}
\mathcal{A}=2\ell_\perp^{d-2}z_*^{d-1}\int_{0}^{\ell/2}\frac{dx}{z(x)^{2(d-1)}}=2\ell_\perp^{d-2}z_*^{d-1}\int_{\epsilon}^{z_*}\frac{dz}{z^{d-1}\sqrt{f(z)[z_*^{2(d-1)}-z^{2(d-1)}]}}
\ ,
\ee
with $z(x)$ is a solution of (\ref{coneq2}). The cutoff scale $\epsilon\ll z_*$ is necessary to render the integral (\ref{onshellA}) finite. On the other hand, the
relation between $z_*$ and $\ell$ can be obtained from
\be\label{ellint}
\frac{\ell}{2}=\int_0^{z_*}\frac{dz}{\sqrt{f(z)[(z_*/z)^{2(d-1)}-1]}} \ .
\ee
The last equation can be written in the following way as a double sum
\begin{align}
\ell=\frac{z_*}{d-1}\sum_{n=0}^\infty \sum_{k=0}^n \frac{\Gamma\left[\frac{1}{2}+n\right]\Gamma \left[\frac{d (n+k+1)-2k}{2 (d-1)}\right]\varepsilon^{n-k} (1-\varepsilon)^k}{
\Gamma[1+n-k]\Gamma[k+1]\Gamma \left[\frac{d (n+k+2)-2k-1}{2 (d-1)}\right]}  \left(\frac{z_*}{z_H}\right)^{n d +k(d-2)} \ .\label{eerc}
\end{align}
And similarly, one can show that
\begin{align}\label{aren}
\mathcal{A}&= \frac{2}{d-2}\left(\frac{\ell_\perp}{\epsilon}\right)^{d-2}+2 \frac{\ell_\perp^{d-2}}{z_*^{d-2}} \left[\frac{\sqrt{\pi } \Gamma \left(-\frac{d-2}{2
(d-1)}\right)}{2 (d-1) \Gamma \left(\frac{1}{2 (d-1)}\right)}\right]\\
&+  \frac{\ell_\perp^{d-2}}{(d-1)z_*^{d-2}} \left[\sum_{n=1}^\infty \sum_{k=0}^n\frac{\Gamma\left[\frac{1}{2}+n\right]\Gamma \left[\frac{d (n+k-1)-2k+2}{2
(d-1)}\right]\varepsilon^{n-k} (1-\varepsilon)^k}{ \Gamma[1+n-k]\Gamma[k+1]\Gamma \left[\frac{d (n+k)-2k+1}{2 (d-1)}\right]}   \left(\frac{z_*}{z_H}\right)^{n d
+k(d-2)}\right].\nonumber
\end{align}
The rest of the procedure, in principle,  is simple. We have to solve equation (\ref{eerc}) for $z_*$ and then we can calculate the area by using equation
(\ref{aren}). Then, entanglement entropy of a rectangular strip can be computed using the relation (\ref{rt}). In practice, however this procedure can be performed
exactly only at zero temperature and chemical potential.

\subsection{Zero temperature and chemical potential}
This case can be solved exactly. At zero temperature and chemical potential $f(z)=1$ and one can integrate (\ref{coneq2}), to obtain
\be\label{embAdS}
x(z)=\frac{\ell}{2}-\frac{z_*}{d}\left(\frac{z}{z_*}\right)^{d}\,\!_2F_1\left[\frac{1}{2},\frac{1}{2}+\frac{1}{2(d-1)};\frac{3}{2}+\frac{1}{2(d-1)};\left(\frac{z}{z_*}\right)^{2(d-1)}\right]
\ .
\ee
Imposing $x(z_*)=0$ in the above, one finds
\be\label{ellzst}
\frac{\ell}{2}=\frac{\sqrt{\pi}\Gamma(\frac{1}{2}+\frac{1}{2(d-1)})}{\Gamma(\frac{1}{2(d-1)})}z_* \ .
\ee
Finally, for $d\geq3$ the area of the extremal surfaces read
\bea
&&
\mathcal{A}=\frac{2\ell_\perp^{d-2}}{d-2}\left[\frac{1}{\epsilon^{d-2}}-\frac{1}{\ell^{d-2}}\left(\frac{\sqrt{\pi}\Gamma(\frac{1}{2}+\frac{1}{2(d-1)})}{\Gamma(\frac{1}{2(d-1)})}\right)^{d-1}\right]+\ldots
\ ,\label{fzv3d}
\eea
where the dots are terms that vanish in the limit $\epsilon\to0$. Therefore the entanglement entropy of the rectangular strip for the boundary theory is given by,
\bea\label{EEAdSpart}
&& S_A=\frac{L^{d-1}}{4
G_N^{(d+1)}}\left(\frac{2}{d-2}\right)\left[\frac{\ell_\perp^{d-2}}{\epsilon^{d-2}}-\frac{\ell_\perp^{d-2}}{\ell^{d-2}}\left(\frac{\sqrt{\pi}\Gamma(\frac{1}{2}+\frac{1}{2(d-1)})}{\Gamma(\frac{1}{2(d-1)})}\right)^{d-1}\right]
\ .
\eea

\subsection{Low effective temperature regime}
\begin{figure}[htbp]
\begin{center}
\includegraphics[height=7cm]{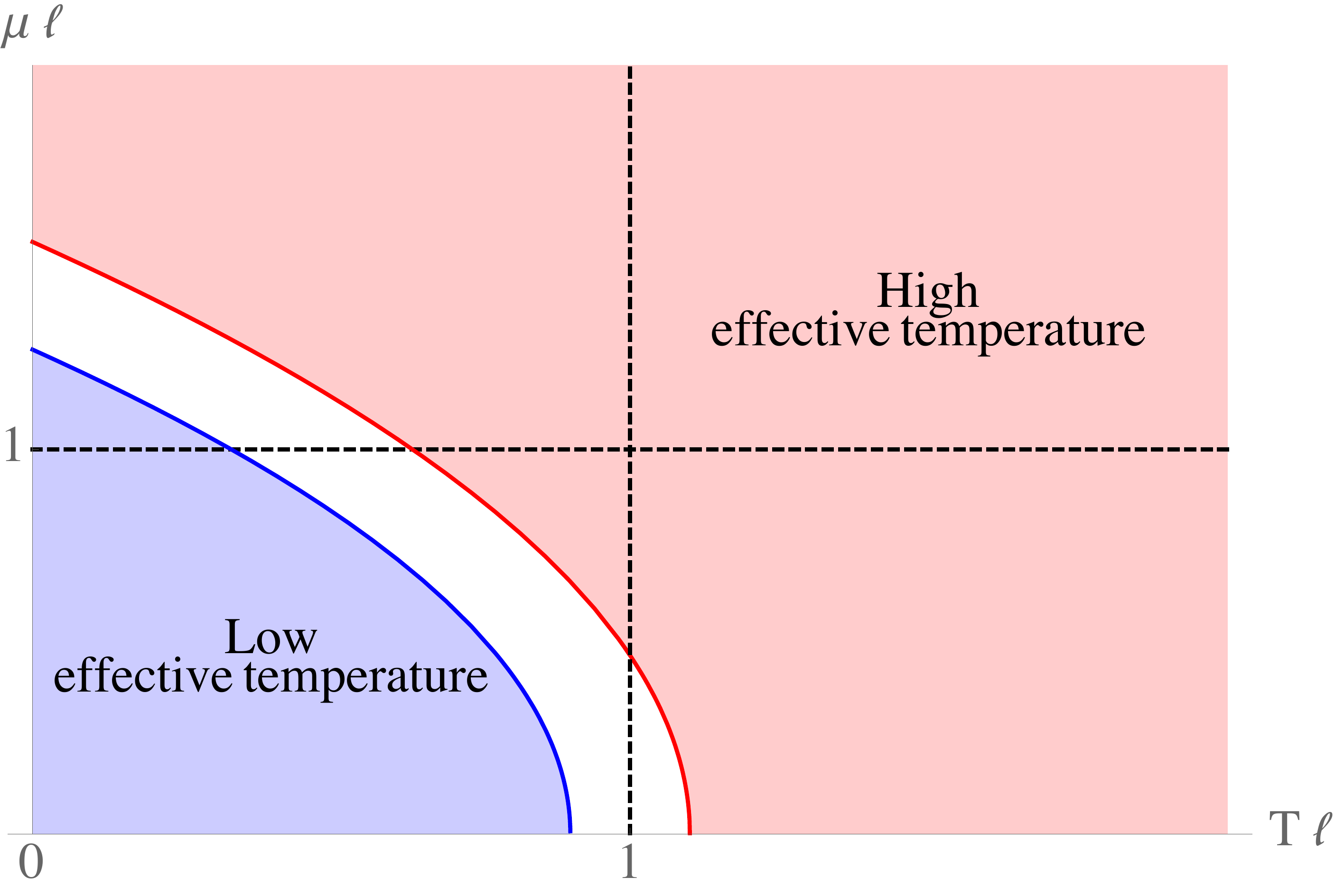}
\caption{For holographic theories, $\T$ completely determines the behavior of entanglement entropy and it is sufficient to only consider (i) high effective temperature regime and (ii) low effective temperature regime.}
\label{regimes2}
\end{center}
\end{figure}

Now we will use the series expansions (\ref{eerc}) and (\ref{aren}) to study extremal surfaces which are restricted to be near the boundary region i.e.
$z_*<<z_H$ (see figure \ref{extsurf}). The leading contributions to the area come from the AdS boundary and hence we should expect the zero temperature entanglement entropy as the dominant
term. The effect of deviation of the bulk geometry from pure AdS is small and  can be computed perturbatively. The equation (\ref{eerc}) can be solved for $z_*$
and at first order in $(\l/z_H)^d$, we obtain
\begin{align}
z_*=\frac{\ell~\Gamma\left[\frac{1}{2(d-1)}\right]}{2 \sqrt{\pi}\Gamma\left[\frac{d}{2(d-1)}\right]} \left[1- \frac{1}{2(d+1)}\frac{2^{\frac{1}{d-1}-d} \Gamma
\left(1+\frac{1}{2 (d-1)}\right) \Gamma \left(\frac{1}{2 (d-1)}\right)^{d+1}}{\pi^{\frac{d+1}{2}}  \Gamma
\left(\frac{1}{2}+\frac{1}{d-1}\right)\Gamma\left(\frac{d}{2(d-1)}\right)^d}\a\left(\frac{\l}{z_H}\right)^d\right.\nonumber\\
\left.+\O\left(\frac{\l}{z_H}\right)^{2(d-1)}\right]\ .
\end{align}
Now using equation (\ref{aren}), at first order in $(\l/z_H)^d$, we get
\begin{align}
\A=\frac{2\ell_\perp^{d-2}}{d-2}\frac{1}{\epsilon^{d-2}}+ \S_0 \left(\frac{\l_\perp}{\l}\right)^{d-2}\left[1+  \varepsilon \S_1\left(\frac{4\pi \T
\l}{d}\right)^d+\O\left(\frac{4\pi \T \l}{d}\right)^{2(d-1)}\right]\ .
\end{align}
Note that $\a \sim \O(1)$ and  numerical constants $\S_0$ and $\S_1$ are given by (\ref{s0}-\ref{s1}). This result can be used to compute the corrections to the entanglement entropy at finite temperature and chemical potential when $\T\l<<1$ (see figure \ref{regimes2}), yielding
\begin{align}\label{result1}
S_A=\frac{L^{d-1}}{4
G_N^{(d+1)}}\left[\frac{2\ell_\perp^{d-2}}{d-2}\frac{1}{\epsilon^{d-2}}+ \S_0 \left(\frac{\l_\perp}{\l}\right)^{d-2}\left\{1+  \varepsilon \S_1\left(\frac{4\pi \T
\l}{d}\right)^d+\O\left(\frac{4\pi \T \l}{d}\right)^{2(d-1)}\right\}\right]\ .
\end{align}
Let us now comment on the above result. First of all entanglement entropy increases with increasing effective temperature. The leading correction term is $\sim \varepsilon T^d$ which is expected from entanglement thermodynamics. Over the last several years thermodynamics of entanglement entropy has attracted a lot of attention \cite{Bhattacharya:2012mi,Allahbakhshi:2013rda}. Variance of entanglement entropy, $\Delta S_A$ and variance of energy $\Delta E_A$ obey a relation which is similar to the first law of thermodynamics $\Delta E_A=T_{ent}\Delta S_A$ where, $\Delta E_A=\mathcal{E} V_A$, and $T_{ent}$ is the so called ``entanglement temperature''. For small intervals, $T_{ent}$ is universal, in the sense that it does not depend on the details of the excitations  \cite{Bhattacharya:2012mi}. From equation (\ref{result1}), we can write
\be
\Delta S_A\approx \frac{L^{d-1}}{4 G_N^{(d+1)}}\S_0 \S_1\varepsilon \left(\frac{\l_\perp}{\l}\right)^{d-2}  \left(\frac{4\pi \T \l}{d}\right)^d\ .
\ee
Similarly from equation (\ref{epsilondef}), we obtain
\be
\Delta E_A=\frac{L^{d-1} (d-1)\l_\perp^{d-2}\l}{16\pi G^{(d+1)}_{N}}\left(\frac{4 \pi  \T}{d}\right)^d\varepsilon\
\ee
which lead us to
\be
T_{ent}=\frac{(d-1)}{4\pi \S_0 \S_1}\frac{1}{\l}\ ,
\ee
which is independent of both $T$ and $\mu$.

Note that the series expansions (\ref{eerc}) and (\ref{aren}) can be used to calculate correction terms at any order.  For example, the sub-sub-leading term goes as $\sim (\varepsilon-1)(\T \l)^{2(d-1)}$. This term is absent for purely thermal case for which $\varepsilon=1$. In that case, the sub-sub-leading term is $\sim (T \l)^{2d}$.

\subsection{High effective temperature regime}
\begin{figure}[htbp]
\begin{center}
  \includegraphics[angle=0,width=0.8\textwidth]{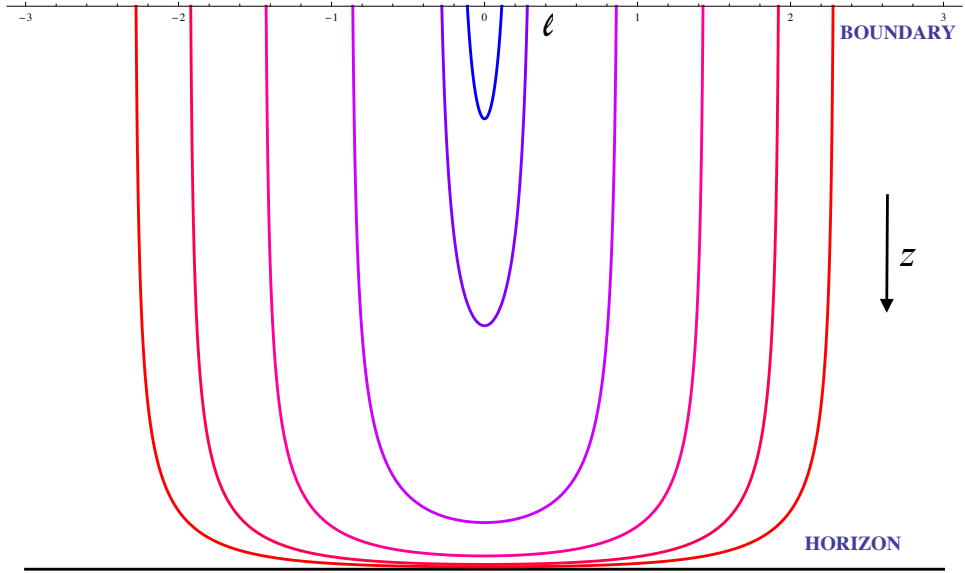}
\end{center}
\caption{\small Behavior of the extremal surfaces for different values of $\l$ with $\T=1$.}
\label{extsurf}
\end{figure}

Let us now consider the limit $z_*\sim z_H$. In this limit extremal surfaces tend to wrap a part of the horizon and the leading contribution comes from this near
horizon part of the surface. It is not very difficult to find out the asymptotic behavior of the extremal surfaces. However, it is more difficult to use the 
systematic expansions (\ref{eerc}-\ref{aren}) directly because $z_*/z_H\sim 1$. Instead one should expand around $z_*/z_H=1$. It is easy to check that both $\l$ and $\A$ diverge in the limit $z_*\rightarrow z_H$.\footnote{This divergence is different from the
UV-divergence of the area $\A$.} Let us first compute the quantity
\begin{align}
\A-\frac{\l_\perp^{d-2}\l}{z_*^{d-1}}=&
2\ell_\perp^{d-2}\int_{\epsilon}^{z_*}\frac{dz}{\sqrt{f(z)[z_*^{2(d-1)}-z^{2(d-1)}]}}\left(\frac{z_*^{d-1}}{z^{d-1}}-\frac{z^{d-1}}{z_*^{d-1}}\right)
\nonumber\\
=& 2\ell_\perp^{d-2}\int_{\epsilon}^{z_*}\frac{dz\sqrt{[z_*^{2(d-1)}-z^{2(d-1)}]}}{z_*^{d-1}z^{d-1}\sqrt{f(z)}}\ .
\end{align}
Now one can show that the last integral does not diverge in the limit $z_*\rightarrow z_H$ (only divergence comes from the boundary $z=\epsilon$). From the last
equation we can write 
\begin{align}
\A-\frac{\l_\perp^{d-2}\l}{z_*^{d-1}}&= \frac{2}{d-2}\left(\frac{\ell_\perp}{\epsilon}\right)^{d-2}+2 \frac{\ell_\perp^{d-2}}{z_*^{d-2}} \left[\frac{\sqrt{\pi }
\Gamma \left(-\frac{d-2}{2 (d-1)}\right)}{2 (d-1) \Gamma \left(\frac{1}{2 (d-1)}\right)}\right]\nonumber\\
&+2\ell_\perp^{d-2}\int_{0}^{z_*}\left(\frac{dz\sqrt{[z_*^{2(d-1)}-z^{2(d-1)}]}}{z_*^{d-1}z^{d-1}\sqrt{f(z)}}-\frac{z_*^{d-1}dz}{z^{d-1}\sqrt{[z_*^{2(d-1)}-z^{2(d-1)}]}}\right)
\ .
\end{align}
In the limit $z_*\rightarrow z_H$, we obtain
\begin{align}
\A\approx & \frac{2}{d-2}\left(\frac{\ell_\perp}{\epsilon}\right)^{d-2}+\frac{\l_\perp^{d-2}\l}{z_H^{d-1}}+ \frac{\ell_\perp^{d-2}}{z_H^{d-2}} N(\varepsilon)\ ,
\end{align}
where $N(\varepsilon)$ is given by
\begin{align}\label{nd}
N(\varepsilon)=2\left[\frac{\sqrt{\pi } \Gamma \left(-\frac{d-2}{2 (d-1)}\right)}{2 (d-1) \Gamma \left(\frac{1}{2
(d-1)}\right)}\right]+2\int_{0}^{1}dx\left(\frac{\sqrt{1-x^{2(d-1)}}}{x^{d-1}\sqrt{f(z_H x)}}-\frac{1}{x^{d-1}\sqrt{1-x^{2(d-1)}}}\right)\ .
\end{align}
This result is going to be very useful for computing the entanglement entropy when the effective temperature $\T$ of the system is very high
compare to $1/\l$. At high effective temperature, the contributions of finite temperature and/or chemical potential to the entanglement entropy become more and more significant and in the dual gravity theory the extremal surface associated with the entangling region approaches the horizon exponentially fast\footnote{See appendix \ref{approach}.} but it always stays finite distance above the horizon (see figure \ref{extsurf}). Therefore, entanglement entropy for $\T\l\gg 1$ (see figure \ref{regimes2}) is given by,
\be\label{result2}
S_A\approx\frac{L^{d-1}}{4 G_N^{(d+1)}} \left[\frac{2}{d-2}\left(\frac{\l_\perp}{\epsilon}\right)^{d-2}+V \left(\frac{4 \pi \T}{d}\right)^{d-1}\left\{1+
\left(\frac{d}{4 \pi \T \l }\right)N(\varepsilon)\right\}\right]\ .
\ee
The higher order corrections to the above results can be easily calculated by extending this procedure in a manner similar to the purely thermal case \cite{Fischler:2012ca}.

\section{Entanglement entropy in different regimes}\label{section4}
In this section we will use the results (\ref{result1}) and (\ref{result2}) to derive analytic expressions for entanglement entropy in different regimes of interest. Before we proceed, let us schematically show the regions in the parameter space $\{T\l, \mu\l\}$ where our analytic expressions are valid:
\be\nonumber
\includegraphics[height=7cm]{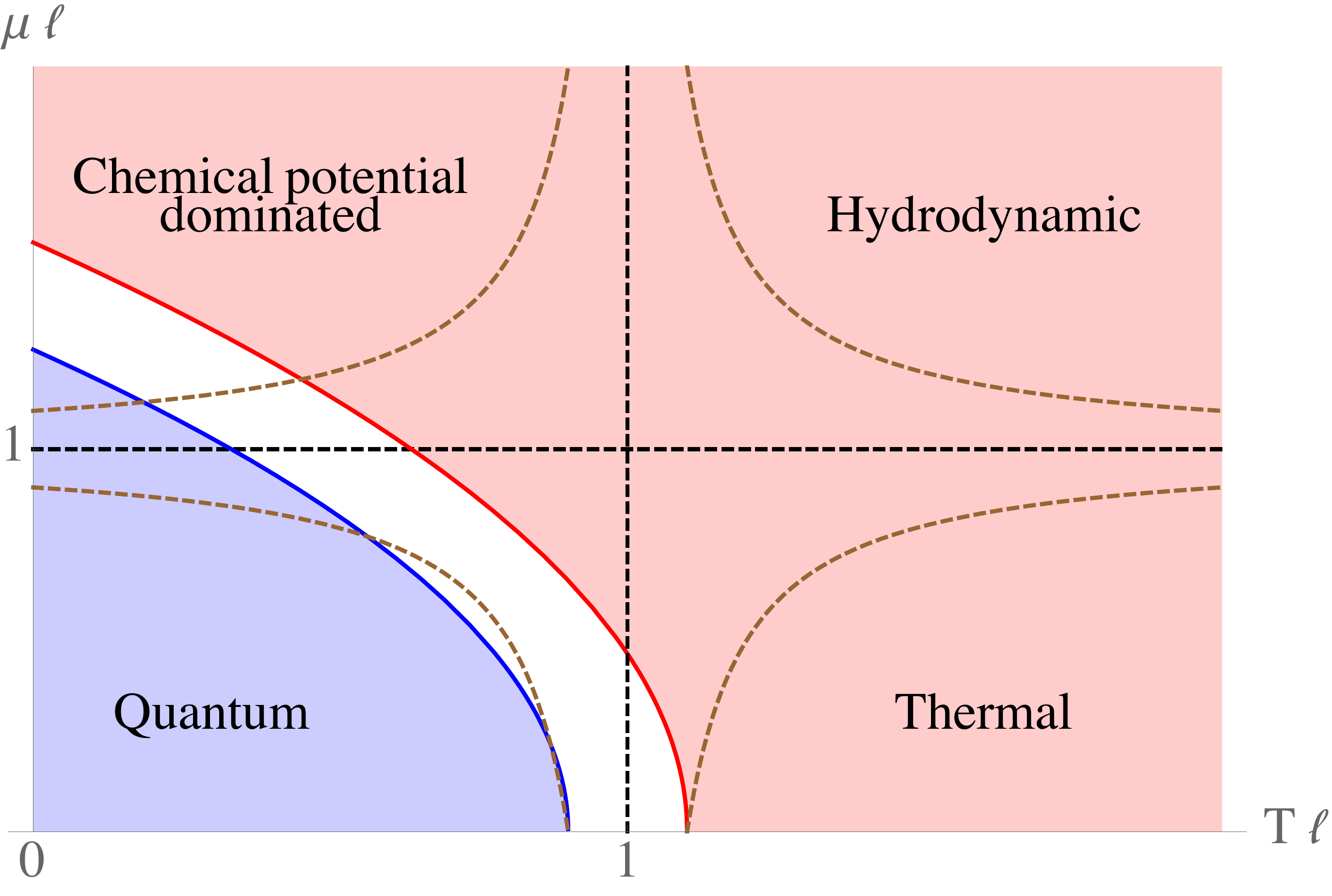}
\label{regimes3}
\ee
The expression  (\ref{result1}) of the entanglement entropy is valid in the blue region, whereas the expression (\ref{result2}) is valid in the red region. As shown in the above figure, most of the thermal, chemical potential dominated and hydrodynamic regimes can be analyzed by using the high effective temperature result, which suggests that there is an effective hydrodynamic description exists for these three regimes.

\subsection{Thermal case: $T\neq 0\ ,\ \mu=0$}
First let us review the finite temperature results of \cite{Fischler:2012ca}. This is a special case with $\varepsilon=1$ and $\T=T$.
\subsubsection{Low temperature limit}
At low temperature ($T\l<<1$), we should expect the zero temperature entanglement entropy as the leading term. In this limit $z_*<< z_H$ and the extremal
surfaces are restricted to be near the boundary region. Finite temperature corrections can be computed perturbatively and using (\ref{result1}), we get
\begin{equation}\label{eel}
S_A=\frac{L^{d-1}}{4 G_N^{(d+1)}}\left[\frac{2}{d-2}\left(\frac{\ell_\perp}{\epsilon}\right)^{d-2}+ \S_0 \left(\frac{\l_\perp}{\l}\right)^{d-2}\left\{1+ \S_1
\left(\frac{4 \pi T \l}{d}\right)^d+ \O\left(\frac{4 \pi T \l}{d}\right)^{2d}\right\}\right]
\end{equation}
where numerical constants $\S_0, \S_1$ are given by equations (\ref{s0}) and (\ref{s1}). Particularly for $d=4$ we have,
\begin{equation}
S_A=\frac{L^{d-1}}{4 G_N^{(d+1)}}\left[\left(\frac{\l_\perp}{\epsilon}\right)^2 -0.32 \left(\frac{\l_\perp}{\l}\right)^{2}\left\{1-(1.764) (\pi T \l)^4+ \O(\pi T
\l)^{8}\right\}\right].
\end{equation}
\subsubsection{High temperature limit}
At high temperature (i.e. $T \l \gg 1$), the extremal surfaces tend to wrap a part of the horizon and the leading contribution comes from the near horizon part
of the surface (\ref{result2}). For the subleading terms, the full bulk geometry contributes and they are more interesting \cite{Fischler:2012ca}. Using
(\ref{result2}) we can reproduce the result of the entanglement entropy of the rectangular strip for the $d$-dimensional boundary theory at high temperature \cite{Fischler:2012ca}
\begin{equation}\label{highee}
S_A\approx\frac{L^{d-1}}{4 G_N^{(d+1)}} \left[\frac{2}{d-2}\left(\frac{\l_\perp}{\epsilon}\right)^{d-2}+V \left(\frac{4 \pi T}{d}\right)^{d-1}\left\{1+
\left(\frac{d}{4 \pi T \l }\right)\S_{high}\right\}\right]
\end{equation}where, $\S_{high}=N(\varepsilon=1)$ is another numerical constant and $V=\l \l_\perp^{d-2}$ is the volume of the rectangular strip.

\subsection{Extremal case: $T=0\ , \ \mu\neq 0$}
The extremal case corresponds to
\be
Q^2=d(d-1)L^2/(d-2)^2z_H^{2(d-1)}
\ee
which is equivalent to $\varepsilon=b$. In the extremal case $T=0$ and the chemical potential is given by
\be
\mu=\frac{1}{z_H}\sqrt{\frac{ab}{2}}=\frac{1}{z_H}\sqrt{\frac{d (d-1)}{(d-2)^2}} \ .
\ee
However, $\T$ is non-zero and is proportional to $\mu$
\be\label{k1}
\T=\frac{\mu d}{2\pi\sqrt{2 a b}}\ .
\ee
Therefore, we expect that the behavior of entanglement entropy at finite chemical potential is going to be somewhat similar to the purely thermal case.
\subsubsection{$\mu \l<<1$ limit}
In the limit $\mu \l<<1$, $z_*<< z_H$ and the leading contributions to the area come from the boundary which is still AdS. Therefore we expect that the vacuum entanglement entropy should be the leading term. Finite chemical potential corrections correspond to the deviation of the bulk geometry from pure
AdS. In the limit $\mu \l<<1$, the extremal surface is restricted to be near the boundary region and hence we can use (\ref{result1}) to compute entanglement entropy of the rectangular strip for the $d$-dimensional boundary theory:
\begin{align}\label{eel}
S_A=&\frac{L^{d-1}}{4 G_N^{(d+1)}}\left[\frac{2}{d-2}\left(\frac{\l_\perp}{\epsilon}\right)^{d-2}\right.\nonumber\\
&\left.+ \S_0 \left(\frac{\l_\perp}{\l}\right)^{d-2}\left\{1+ \frac{2(d-1)\S_1}{d-2}
\left(\frac{(d-2)\mu\l}{\sqrt{d(d-1)}}\right)^d+\O\left(\mu\l\right)^{2(d-1)}\right\}\right]\ ,
\end{align}
where, numerical constants $\S_0$ and  $\S_1$ are given by equations (\ref{s0}) and (\ref{s1}). Note that both $\S_0$ and $\S_1$ are negative. Let us now compute the change in the entanglement entropy because of the chemical potential
\be
\Delta S_A(\mu)=S_A(\mu)-S_A(\mu=0)=\left(\frac{L^{d-1}}{4 G_N^{(d+1)}}\right)\frac{2(d-1)\S_0\S_1}{d-2} \left(\frac{(d-2)}{\sqrt{d(d-1)}}\right)^d \mu^d
\l_\perp^{d-2}\l^2 \ .
\ee
Since $\S_0\S_1>0$, chemical potential increases the entanglement entropy. Let us also compute the quantity $\Delta S_A(T)$, which is the correction to the
entanglement entropy because of the temperature. In the low temperature limit we have
\be
\Delta S_A(T)=S_A(T)-S_A(T=0)=\left(\frac{L^{d-1}}{4 G_N^{(d+1)}}\right)\S_0\S_1 \left(\frac{4\pi}{d}\right)^d T^d \l_\perp^{d-2}\l^2 \ .
\ee
Therefore, in the limit $\mu \l<<1$,  $\Delta S_A(\mu)$ has exactly the same functional form as $\Delta S_A(T=\T)$, with $\T$ given by (\ref{k1}). They only
differ by an overall factor of $a\sim \O(1)$.

\subsubsection{$\mu \l>>1$ limit}
For large chemical potential (i.e. $\mu \l \gg 1$), $z_* \sim z_H$ and the leading contribution comes from the near horizon part of the surface.\footnote{At $T=0$ but finite $\mu$, the extremal surface approaches the horizon at a power law rate in the limit $\mu\l\gg1$ (see appendix \ref{approach}).} Whereas,  the full bulk geometry contributes to the subleading terms. Using the result (\ref{result2}) we can write
\begin{align}\label{highmu}
S_A\approx\frac{L^{d-1}}{4 G_N^{(d+1)}}\left[
\frac{2}{d-2}\left(\frac{\ell_\perp}{\epsilon}\right)^{d-2}+\left(\frac{(d-2)}{\sqrt{d(d-1)}}\right)^{d-1}V\mu^{d-1}\right.~~~~~~~~~~~\nonumber\\
\left.~~~~~~~+A_d\left(\frac{(d-2)}{\sqrt{d(d-1)}}\right)^{d-2}\l_\perp^{d-2} \mu^{d-2}\right]\ .
\end{align}
Where, $V=\l_\perp^{d-2}\l$ is the volume of the strip and  $A_d$ is given by $A_d=N(\varepsilon=a)$. It is easy to check that $A_d$ is finite and
\begin{align}
A_3=0.106\ , \qquad A_4=0.878\ , \qquad A_5=1.063\ , \qquad... \nonumber
\end{align}

\begin{figure}[tb!]
\centering
\includegraphics[width=0.6\textwidth]{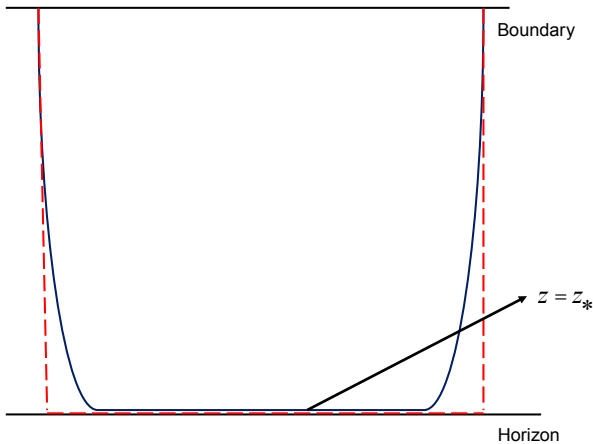}
\caption{For $\T\l>>1$, the actual geodesic (solid blue line) can be approximated by the dashed red line curve that consists of $x=-l/2, z=z_H, x=l/2$.}
\label{approxgeodesic}
\end{figure}
The divergent part of the entanglement entropy is independent of $\mu$ and thus it does not contain any new information. The leading finite piece in equation
(\ref{highmu}) is proportional to the volume of the rectangular strip.  The extrinsic nature of the  leading term at large $\mu$ can be understood very easily by
looking at the extremal surface for $\l/z_H\gg 1$. In this limit, the extremal surface tends to wrap a part of the horizon and the actual U-shaped surface can be
approximated by a surface that consists of $x=-\l/2, z=z_H, x=\l/2$ (see figure \ref{approxgeodesic}). At large $\mu$ limit, the most dominant contribution to the area of the extremal surface
comes from the near horizon part which can be guessed from this approximate surface. The area of the near horizon part of the approximate surface is $A\sim
V/z_H^{d-1} $. Therefore, it is expected that the leading term goes as $S_A\sim V \mu^{d-1}$. On the other hand, the other term $\sim (\l_\perp \mu)^{d-2}$ is
more interesting. One can guess the functional form of this term from the approximate surface. However, the value of the numerical constant $A_d$ obtained from
the approximate surface is inaccurate.

Let us now compare the entanglement entropy at high temperature $(T\l>>1)$ and the entanglement entropy for large chemical potential $(\mu \l>>1)$. In both cases
the leading finite part is proportional to the volume $V$ of the strip and
\be
S_A(T=0,\mu)=S_A(T=\T,\mu=0)+\O\left(\frac{1}{\mu \l} \right)\ .
\ee

The calculation of the entanglement entropy of an infinite rectangular strip suggests that for holographic theories the general form of the finite part of the entanglement entropy for
large $\mu$ does not particularly depend on the shape. One expects that the finite part of the entanglement entropy of a region $A$ for a $d-$dimensional ($d\ge 3$)
boundary theory is given by
\begin{align}
S_{A;finite}=c_0 &\left[ \mu^{d-1}\text{Volume}(A)+c_1 ~\mu^{d-2}\text{Area}(\partial A)\right]+ \text{sub-leading terms},
\end{align}
provided  $\mu\gg1/l$, where $l$ is the smallest length scale of the region $A$. $c_0$ is a constant that depends on the particular theory and $c_1$ is a constant
that depends on the shape of the region $A$.

\subsection{Near thermal case: $T>>\mu$}
In this limit, $\varepsilon(T,\mu)$ and $\T(T,\mu)$ are given by,
\begin{align}
&\T=T\left(1+\frac{d(d-2)^2}{16\pi^2(d-1)}\left(\frac{\mu}{T}\right)^2+\O\left(\frac{\mu}{T}\right)^4\right)\ ,\\
&\a=1+\frac{d^2(d-2)}{16\pi^2(d-1)}\left(\frac{\mu}{T}\right)^2+\O\left(\frac{\mu}{T}\right)^4\ .
\end{align}
Presence of the chemical potential increases the effective temperature of the system.

\subsubsection{$\mu \l << T \l<<1$ limit:}
Using (\ref{result1}), in the limit $\mu \l << T \l<<1$, we obtain
\begin{equation}\label{eel}
S_A=\frac{L^{d-1}}{4 G_N^{(d+1)}}\left[\frac{2}{d-2}\left(\frac{\ell_\perp}{\epsilon}\right)^{d-2}+ \S_0 \left(\frac{\l_\perp}{\l}\right)^{d-2}\left\{1+\a \S_1
\left(\frac{4 \pi \T \l}{d}\right)^d+\O\left(\T\l\right)^{2(d-1)}\right\}\right]
\end{equation}
where $\S_0$ and $\S_1$ are given in equations (\ref{s0}, \ref{s1}). Therefore, in this limit  $\mu \l << T \l<<1$, in the leading order, we obtain
\be
\Delta S_A(T,\mu)=S_A(T,\mu)-S_A(T,\mu=0)=\left(\frac{L^{d-1}(d-2)}{4 G_N^{(d+1)}}\right)\S_0\S_1 \left(\frac{4\pi}{d}\right)^{d-2} T^{d-2}\mu^2
\l_\perp^{d-2}\l^2 \ .
\ee
Note that $\Delta S_A(T,\mu)>0$.

\subsubsection{$T>>\mu$ and $T\l>>1$:}
Let us now look at the other limit $T>>\mu$ and $T\l>>1$. In this limit, we can again use (\ref{result2}) which leads to
\begin{equation}\label{extremalcase}
S_A(T,\mu)\approx\frac{L^{d-1}}{4 G_N^{(d+1)}} \left[\frac{2}{d-2}\left(\frac{\l_\perp}{\epsilon}\right)^{d-2}+V \left(\frac{4 \pi
\T}{d}\right)^{d-1}+\l_\perp^{d-2}\left(\frac{4 \pi \T}{d}\right)^{d-2}\gamma_d\left(\frac{\mu}{T}\right)\right]
\end{equation}
where $V=\l \l_\perp^{d-2}$ is the volume of the rectangular strip. The function $\gamma_d$ is given by,
\begin{align}
\gamma_d\left(\frac{\mu}{T}\right)=N(1)+\frac{d^2(d-2)}{16\pi^2(d-1)}\left(\frac{\mu}{T}\right)^2\int_{0}^{1}dx\left(\frac{x\sqrt{1-x^{2(d-1)}}}{\sqrt{1-x^d}}\right)\left(\frac{1-x^{d-2}}{1-x^d}\right)+\O\left(\frac{\mu}{T}\right)^4\
.
\end{align}
Where, the function $N(\varepsilon)$ is defined by (\ref{nd}). Therefore, in the limit $T>>\mu$ and $T\l>>1$ at the leading order, we obtain
\be
\Delta S_A(T,\mu)=S_A(T,\mu)-S_A(T,\mu=0)=\left(\frac{L^{d-1}(d-2)^2}{4 G_N^{(d+1)}}\right)\left(\frac{4\pi}{d}\right)^{d-3} T^{d-3}\mu^2 V\ .
\ee
The correction to the part of the entanglement entropy which is proportional to area is more interesting since that corresponds to the actual entanglement. In
general it has the following form
\begin{align}
\Delta S_A(T,\mu)= c_0 T^{d-3}\mu^2 \l \l_\perp^{d-2}+c_1 ~T^{d-4}\mu^2\l_\perp^{d-2}+... \ ,
\end{align}
where, $c_0$ and $c_1$ can be found from (\ref{extremalcase}) by performing a series expansion in $\mu/T$.
Later we will show that mutual information subtracts out the part of the entanglement entropy which is proportional to volume and hence mutual information  measures the actual quantum entanglement between two sub-regions.
\subsection{Near extremal case: $T<<\mu$}
In the near extremal limit:
\be
\varepsilon \approx a-\frac{2b\pi \sqrt{2ab}}{d}\left(\frac{T}{\mu}\right)\ \qquad \text{and}\qquad \T\approx \frac{1}{2}\left(\frac{\mu d}{\pi\sqrt{2ab}}+T\right)\ .
\ee
\subsubsection{$1>>\mu \l>> T \l$ limit}
In the limit $1>>\mu \l>> T \l$, using (\ref{result1}), we obtain
\begin{align}
S_A=&\frac{L^{d-1}}{4 G_N^{(d+1)}}\left[\frac{2}{d-2}\left(\frac{\l_\perp}{\epsilon}\right)^{d-2}\right.\nonumber\\
&\left.+ \S_0 \left(\frac{\l_\perp}{\l}\right)^{d-2}\left\{1+\varepsilon\S_1 \left(\frac{4\pi \T\l}{d}\right)^d+\O\left(\T\l\right)^{2(d-1)}\right\}\right] \ ,
\end{align}
Therefore, in this limit  $1>>\mu \l>> T \l$, in the leading order, we obtain
\be
\Delta S_A(T,\mu)=S_A(T,\mu)-S_A(T=0,\mu)=\left(\frac{L^{d-1}}{4 G_N^{(d+1)}}\right) \frac{2\pi \S_0\S_1}{(2ab)^{\frac{d-1}{2}}} T\mu^{d-1} \l_\perp^{d-2}\l^2 \
.
\ee
One can check that $\Delta S_A(T,\mu)>0$.

\subsubsection{$T<<\mu$ and $\mu \l>>1$}
In the limit $\mu \l>>1$, we obtain,
\begin{align}\label{extremalcase1}
S_A\approx&\frac{L^{d-1}}{4 G_N^{(d+1)}} \left[\frac{2}{d-2}\left(\frac{\l_\perp}{\epsilon}\right)^{d-2}\right.\\
&~~~~~~~~~~~~~~\left.+V \left(\frac{4 \pi \T}{d}\right)^{d-1}\left\{1+ \left(\frac{d}{4 \pi \T \l }\right)\left(N_0+N_1 (a-\varepsilon)
+\O\left(\frac{T}{\mu}\right)^2\right)\right\}\right]\nonumber
\end{align}
where $N_0$ and $N_1$ are numerical constants given in appendix \ref{factors}. Therefore, in the limit $T<<\mu$ and $\mu \l>>1$ at the leading order, we obtain
\be
\Delta S_A(T,\mu)=S_A(T,\mu)-S_A(\mu,T=0)=\left(\frac{L^{d-1}\pi(d-1)}{2d G_N^{(d+1)}}\right)\left(\frac{2}{ab}\right)^{\frac{d-2}{2}} \mu^{d-2}T V\ .
\ee
The leading correction is linear in $T$. Part of the entanglement entropy which is proportional to area also receives corrections and in general it has the
following form
\begin{align}
\Delta S_A(T,\mu)=c_0 \mu^{d-2}T \l \l_\perp^{d-2}+c_1 ~\mu^{d-3}T\l_\perp^{d-2}+...\ ,
\end{align}
where, $c_0$ and $c_1$ can be found from (\ref{extremalcase1}) by performing a series expansion in $T/\mu$. In the next section, we will show how mutual information subtracts out the part of the entanglement entropy which is proportional to volume and actually measures the  quantum entanglement between two sub-regions.
\section{Mutual information and effective temperature}\label{sectionmi}
The entanglement entropy of a spatial region in a local field theory is UV-divergent. Only local physics contributes to the UV-divergent piece, however, the finite part contains information about the long range entanglement. Mutual information is a quantity that is derived from
entanglement entropy.
\begin{figure}[!]
\centering
\includegraphics[width=0.8\textwidth]{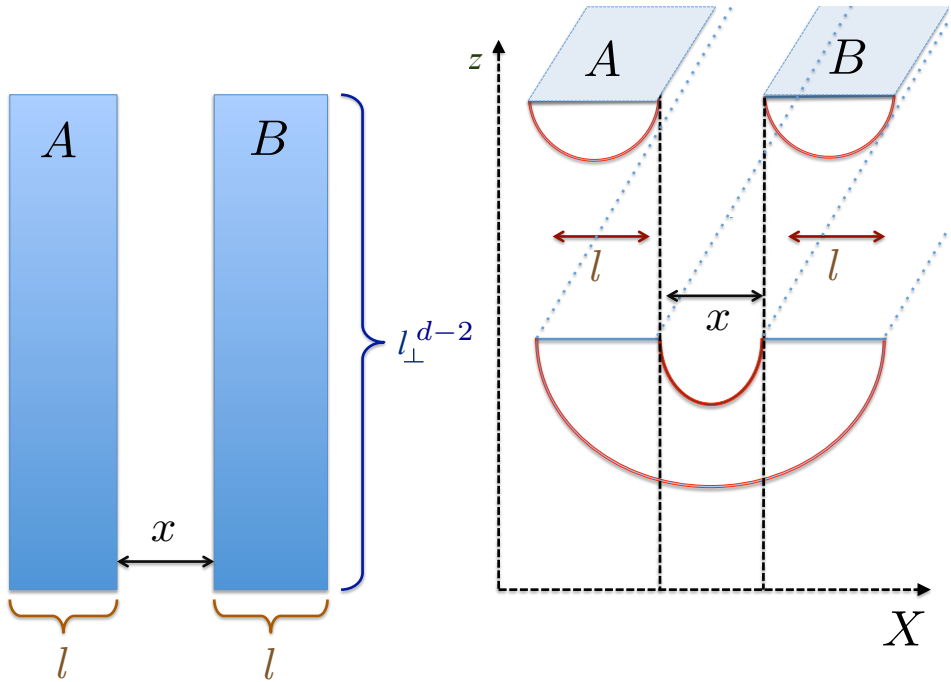}
\caption{The two disjoint sub-systems $A$ and $B$, each of length $l$ along $X$-direction and separated by a distance $x$. The schematic diagram on the right
shows the possible candidates for minimal area surfaces which is relevant for computing $S_{A\cup B}$. See \cite{Fischler:2012uv} for a detailed discussion.}
\label{shape}
\end{figure}
Mutual information between two disjoint sub-systems A and B is defined as
\be
I(A,B)=S_A+S_B- S_{A\cup B}
\ee
where $S_A$, $S_B$ and $S_{A\cup B}$ denote entanglement entropy of the region $A$, $B$ and $A\cup B$ respectively with the rest of the system (see
fig.~\ref{shape} for an example). Mutual information is a UV-finite quantity and hence it does not depend on regularization scheme. Moreover, as showed in
\cite{PhysRevLett.100.070502}, given an operator $\O_A$ in the region $A$ and $\O_B$ in the region $B$, mutual information sets an upper bound
\begin{eqnarray} \label{mi1}
I(A, B) \ge \frac{\left(\langle \O_A \O_B \rangle - \langle \O_A \rangle \langle \O_B \rangle \right)^2}{2 \langle \O_A^2 \rangle \langle \O_B^2 \rangle}
\end{eqnarray}
and thus measures the total correlation between the two sub-systems: including both classical and quantum correlations. Mutual information has several advantages over entanglement entropy. In the context of holography, mutual information has been used extensively to study various aspects of quantum entanglement, for example see  \cite{Headrick:2010zt,Tonni:2010pv,Fischler:2012uv,MolinaVilaplana:2011xt,Fischler:2013gsa,Li:2013sia,Andrade:2013rra,Asplund:2013zba,Agon:2015mja,Hosseini:2015gua,Mozaffar:2015xue}.

In this section we will compute  mutual information between two ``rectangular strips" each of width $\l$ separated by a distance $x$ along the
$x^2$-direction for $d\ge 3$ by using our analytic results of entanglement entropy. For a visual rendition of the set-up, see fig.~\ref{shape}. It is shown schematically in fig.~\ref{shape} that for the entanglement entropy of the region $A\cup B$, we have two possible candidates for the corresponding minimal area surface. As a consequence, mutual information undergoes an interesting entanglement/disentanglement ``phase transition" \cite{Headrick:2010zt,Fischler:2012uv}. In this section, we will show that in the presence of chemical potential this phase transition has interesting features.

\subsection{Zero temperature and chemical potential}
Let us first define,
\be
c_d=\frac{L^{d-1}}{4 G_N^{(d+1)}}\ .
\ee
In $d \ge 3$, at zero temperature and chemical potential, the mutual information is given by,
\begin{align}
I(A, B)|_0 &=c_d\S_0 \l_\perp^{d-2}\left[\frac{2}{\l^{d-2}}-\frac{1}{x^{d-2}}- \frac{1}{(2\l+x)^{d-2}}\right]\ , \qquad \frac{x}{\l}< \alpha_d
\nonumber \\
&= 0\ , \qquad \frac{x}{\l}\ge \alpha_d\ .
\end{align}
Recall that $\S_0$ is a negative constant. $\alpha_d$ is a monotonically increasing function of spacetime dimension $d$.\footnote{For numerical values of
$\alpha_d$, see \cite{Fischler:2012uv}.}

\subsection{Finite temperature and chemical potential}
Now we analytically compute mutual information  for $d\ge 3$, in different limits as a function of $x$, $\l$, $\T$ and $\varepsilon$. Let us rewrite the
definitions of $\T$ and $\varepsilon$ for convenience,
\begin{align}
&\T(T,\mu)=\frac{T}{2}\left[1+\sqrt{1+\frac{d^2}{2\pi^2 a b}\left(\frac{\mu^2}{T^2}\right)}\right]\ , \\
&\varepsilon(T,\mu)=a-\frac{b T}{\T(T,\mu)}\ .
\end{align}

\subsubsection{Low effective temperature: $\T<<\frac{1}{\l},\frac{1}{x}$}
In the limit $\T<<\frac{1}{\l},\frac{1}{x}$, mutual information, when it is non-zero, can be computed from (\ref{result1})
\be
I(A, B)=I(A, B)|_0-2c_d\left(\frac{4\pi}{d}\right)^d \S_0 \S_1\varepsilon \l_\perp^{d-2}\T^d (l+x)^2\ .
\ee
The correction term is negative. It was shown in \cite{Fischler:2012uv} that introduction of finite temperature decreases mutual information. The last equation shows that the leading correction term $\sim\varepsilon \T^d$. Introduction of
chemical potential increases both the effective temperature and the energy parameter $\varepsilon$ and hence further decreases the mutual information.

\subsubsection{Intermediate effective temperature: $\frac{1}{\l}<<\T<<\frac{1}{x}$}
In the limit $\frac{1}{\l}<<\T<<\frac{1}{x}$, mutual information can be calculated using (\ref{result1}, \ref{result2}). When, mutual information is non-zero, it
is given by,
\begin{align}\label{mi1}
I(A, B)=c_d \l_\perp^{d-2}\T^{d-2}\left[ -\frac{\S_0}{(\T x)^{d-2}}+\left(\frac{4\pi}{d}\right)^{d-2}N(\varepsilon)-\left(\frac{4\pi}{d}\right)^{d-1}\T
x\right.~~~~~\nonumber\\
\left. -\S_0\S_1\left(\frac{4\pi}{d}\right)^d\varepsilon\T^2 x^2\right]\ ,
\end{align}
where $N(\varepsilon)$ is given by (\ref{nd}). Few comments are in order. First note that the above expression is independent on $\l$. This has an interesting
consequence that we will discuss later in the paper. Above expression also shows that there is an upper bound on $x\T$ above which mutual information is always
zero. This is similar to the behavior of mutual information for the purely thermal case \cite{Fischler:2012uv}.

\subsubsection{High effective temperature: $\T>>\frac{1}{x}$}
In the limit $\T>>\frac{1}{x}$ mutual information is identically zero because the minimal surface corresponding to $A\cup B$ becomes disconnected. Hence,
$S(A\cup B)=S(A)+S(B)$ and   subsystems $A$ and $B$ are completely disentangled.

\subsection{$x\rightarrow 0$ limit and entanglement}
Behavior of the entanglement entropy at high temperature has been studied in details in \cite{Fischler:2012ca}. In the limit $\l T>>1$, it was found that
\be\label{x1}
S(A)=S_{div}+S_{th}+S_{ent}+S_{corr}\ ,
\ee
where, $S_{div}$ is the divergent part of the entanglement entropy which follows an area law. $S_{th}$ is proportional to volume of the region A and it is
just the thermal entropy contribution of the entanglement entropy. $S_{ent}$ is the subleading term which is proportional to the area of A and it measures the
actual quantum entanglement. $S_{corr}$ denotes the correction terms that are exponentially suppressed. We have seen in the previous section that in the presence
of chemical potential $\mu$, entanglement entropy at high effective temperature $\T \l>>1$ also has the same form as (\ref{x1}), where $S_{th}\sim \T^{d-1}V$
is the thermodynamic entropy and $S_{ent}\sim \T^{d-2}A$ is proportional to the area and hence measures actual entanglement.

In \cite{Fischler:2012uv}, it was observed that in the limit $x\rightarrow 0$, at the leading order mutual information coincides with the thermal-part-subtracted
entanglement entropy and hence measures the actual quantum entanglement. From equation (\ref{mi1}), it is clear that the same is true even in the presence of
chemical potential. In particular,
\be
I(A, B)|_{x\rightarrow 0}=I_{div}+ S_{ent}+I_{corr}\ .
\ee
$I_{div}$ has the same structure as $S_{div}$ and hence follow an area law. The leading finite part is exactly $S_{ent}$,  the actual quantum entanglement part of the entanglement entropy and
it also follows an area law. This is a very non-trivial property of mutual information and we believe this should be studied in more details.

\subsection{Phase transition of mutual information}
\begin{figure}[h]
\begin{center}
\unitlength = 1mm

\subfigure[ ]{
\includegraphics[width=85mm]{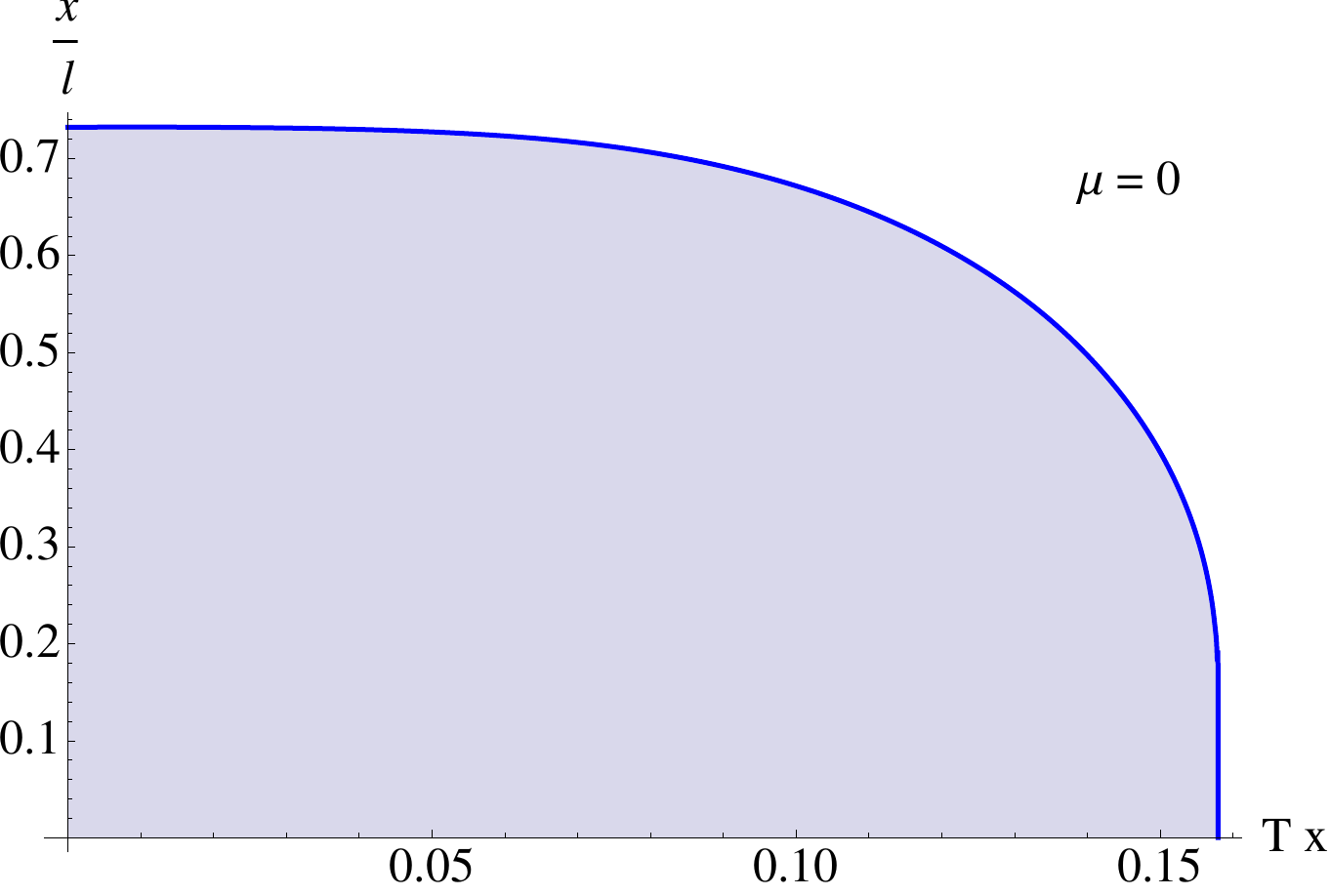}
}
\subfigure[ ]{
\includegraphics[width=85mm]{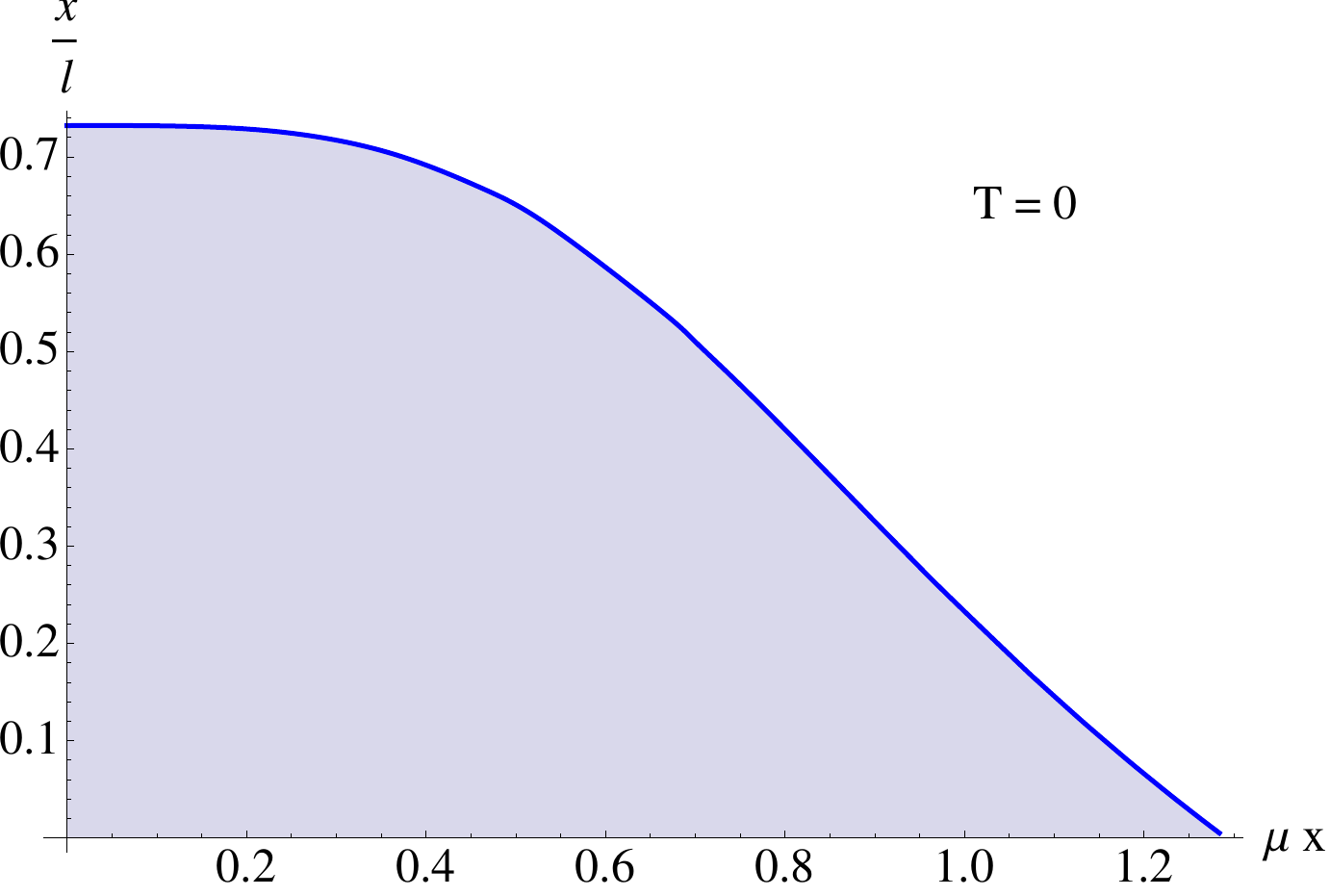}
 }
\caption{\small Phase transition of mutual information: 2-dimensional parameter space for $(3+1)-$dimensional boundary theory at (a) finite temperature but
$\mu=0$ and (b) finite chemical potential but $T=0$. Mutual information is non-zero only in the blue shaded region. } \label{pt1}
\end{center}
\end{figure}
\begin{figure}[h]
\centering
\includegraphics[width=0.8\textwidth]{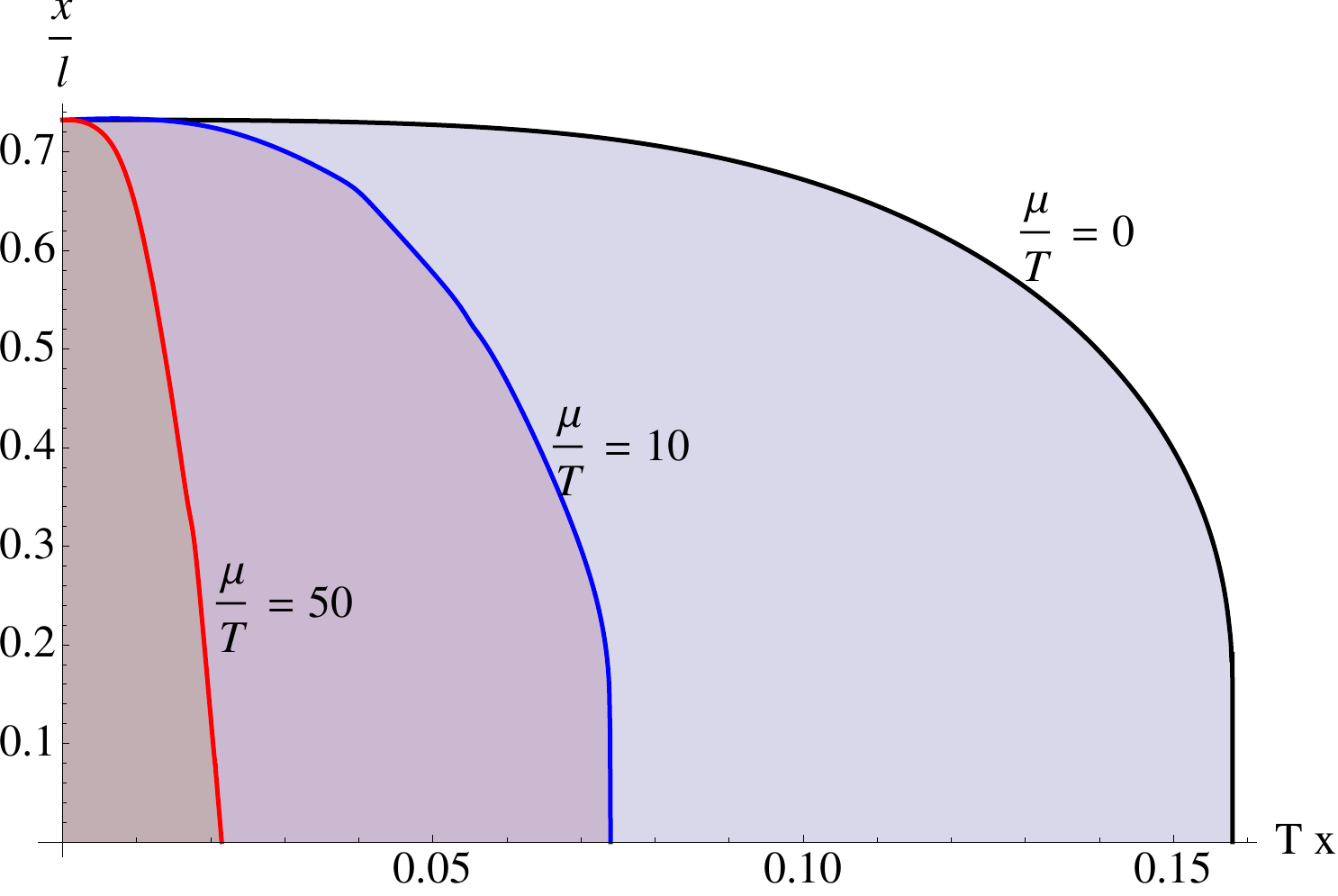}
\caption{Entanglement/disentanglement transition of mutual information at finite temperature and chemical potential: 2-dimensional parameter space for the
(3+1)-dimensional boundary theory. The mutual informational is non-zero only in the shaded region. Mutual information is nonzero for the region below the black
curve for $\mu/T=0$ case ($\varepsilon=1$), below the blue curve for $\mu/T=10$ case ($\varepsilon=2.17$) and below the red curve for $\mu/T=50$ case ($\varepsilon=2.79$). Clearly entanglement between the strips decreases
with increasing chemical potential.}
\label{transplot}
\end{figure}
\begin{figure}[h]
\centering
\includegraphics[width=0.8\textwidth]{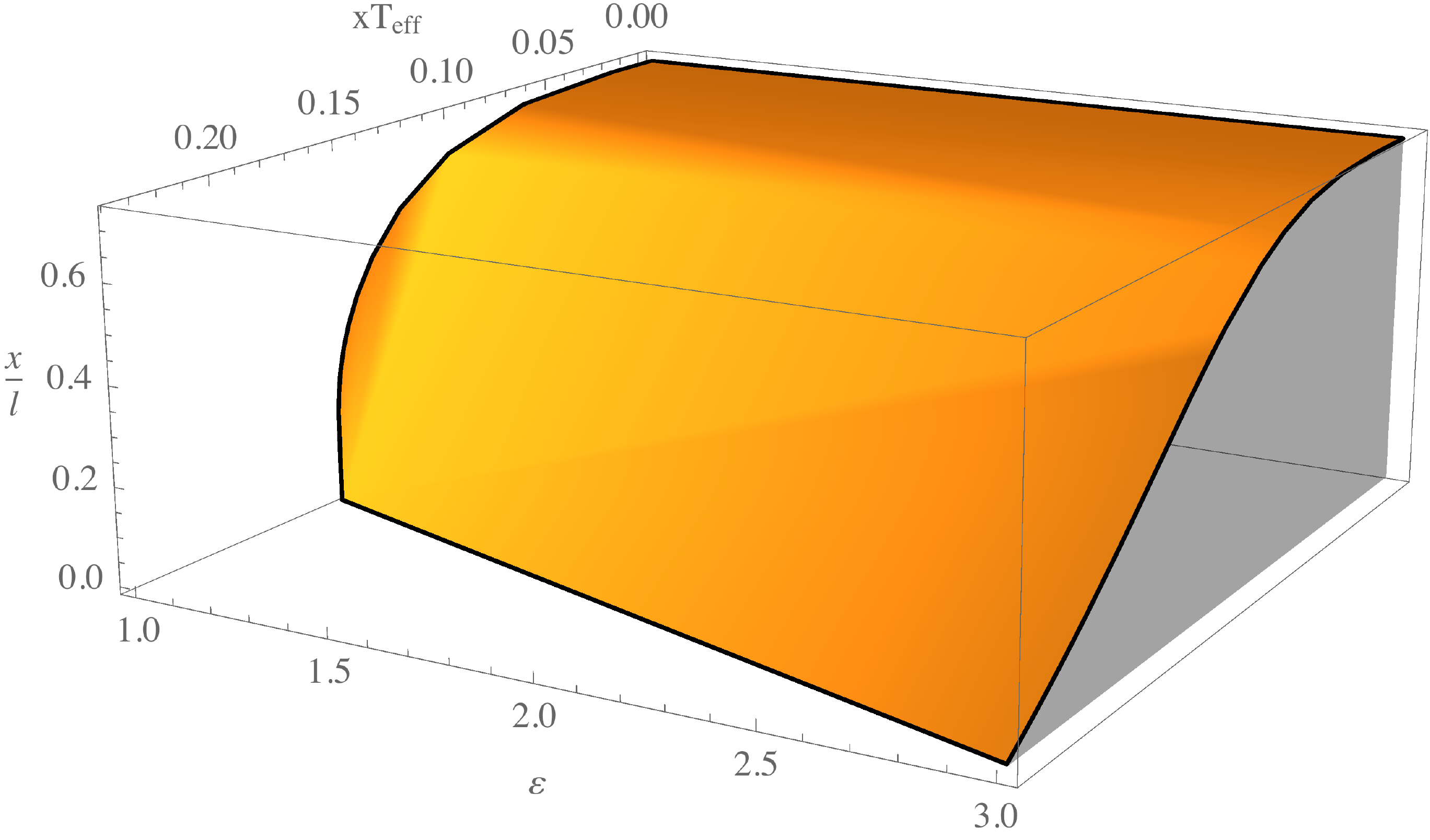}
\caption{Entanglement/disentanglement transition of mutual information at finite temperature and chemical potential: 3-dimensional parameter space $\{\varepsilon, x\T , x/\l\}$ for the
(3+1)-dimensional boundary theory. The mutual informational is non-zero only in the shaded region below the surface. In this case, the energy parameter $1\le \varepsilon \le 3$ saturates the lower bound for the purely thermal case and saturates the upper bound for the zero temperature case.}
\label{transplot3d}
\end{figure}

It is well known that mutual information undergoes an entanglement/disentanglement transition for large N gauge theories which have holographic dual descriptions
\cite{Headrick:2010zt,Fischler:2012uv,MolinaVilaplana:2011xt}. Let us now study this phase transition in the presence of chemical potential. From figure
(\ref{pt1}) it is clear that at finite chemical potential mutual information undergoes an entanglement/disentanglement transition which is similar to the phase
transition at finite temperature but zero chemical potential. In particular, there is an upper bound on $\mu x$ above which mutual information is always zero. In
general,
\begin{align}
&I(A, B) \neq 0\ , \qquad \mu x< \alpha_d\left(\frac{x}{\l}\right) \nonumber \\
&I(A, B) = 0\ , \qquad \mu x\ge \alpha_d\left(\frac{x}{\l}\right)
\end{align}
where $\alpha_d\left(\frac{x}{\l}\right)$ is some function which is shown in figure \ref{pt1}(b) for $d=4$. At finite temperature and chemical potential, this transition has a richer structure which is shown in figures \ref{transplot} and \ref{transplot3d}. From the figure it is clear
that entanglement between region A and B decreases with increasing temperature. At a fixed temperature, the entanglement also decreases with the increase of
chemical potential. This is consistent with our analytic results which also show that the mutual information decreases with the increase of the effective
temperature.

We can also study the entanglement/disentanglement transition of mutual information  as a function of $\T$ and $\varepsilon$. In figure \ref{transplot3d}, we have plotted the 3-dimensional ``phase diagram" of this transition in the parameter space $\{\varepsilon, x\T , x/\l\}$. Interestingly, in the limit $l\rightarrow \infty$, for all the states of the system with the same number of microstates, entanglement between two sub-regions increases with increasing energy parameter $\varepsilon$, i.e. with increasing $\mu/T$.

It is important to note that this sharp phase transition of mutual information is a consequence of the large-$N$ limit. At finite $N$, $1/N$-corrections to the
holographic entanglement entropy formula become important and it is expected from (\ref{mi1}) that mutual information should not vanish identically
\cite{Headrick:2010zt,Fischler:2012uv}. Recently it has been shown explicitly that the leading $1/N$-correction to mutual information is finite for large distances \cite{Agon:2015ftl} and hence the sharp phase transition disappears.

\section{Conclusions}\label{conclusions}

There are numerous examples of strongly coupled systems in nature, ranging from Quantum chromodynamics to various condensed matter systems. In this paper, we have studied the behavior of  entanglement entropy and mutual information at finite temperature and chemical potential in a class of strongly coupled large-N gauge theories in $d$-dimensions that are dual to AdS-RN$_{d+1}$ spacetimes for $d\ge 3$, using the Ryu-Takayanagi formula. The effective temperature $\T$ controls  the nature of the extremal surfaces and hence determines the behavior of entanglement entropy and mutual information in different regimes. The physics is qualitatively very similar to the purely thermal case as studied in  \cite{Fischler:2012ca}, if we replace temperature by the effective temperature. The energy parameter $\varepsilon$ does not play any significant role in determining the behavior of entanglement entropy other than contributing to the coefficients of certain terms.

We have studied properties of mutual information between two disjoint regions for a conformal field theory in $d-$dimensions ($d\ge 3$) at finite temperature and chemical potential. We have shown that mutual information indeed has several advantages over entanglement entropy. In particular, at high effective temperature, entanglement entropy is dominated by the thermodynamic entropy, however, mutual information subtracts out this contribution from the entanglement entropy and measures the actual quantum entanglement. We have  also shown that for these theories mutual information undergoes a transition which is similar to the transition studied in \cite{Fischler:2012uv}. However, at finite chemical potential the three-dimensional ``phase diagram" as shown in figure \ref{transplot3d}, has some particular features. We expect that these features are universal  for quantum field theories, both relativistic and non-relativistic, which have holographic dual descriptions.

There are a few possibilities for future work. One straightforward idea is to consider the $(1+1)-$dimensional case, which has been excluded from this paper due to some technicalities (see appendix \ref{AdSRN3} for details). Even so, we believe that similar analysis can be done in this case. One could also adapt the same machinery to study entanglement entropy in global AdS-RN backgrounds. The interplay between finite volume effects and charge is known to lead to an extremely rich phase structure of the entanglement entropy, including Hawking-Page and Van-der-Waals like transitions \cite{Johnson:2013dka,Caceres:2015vsa,Nguyen:2015wfa}. Another interesting exercise would be to study entanglement entropy in charged black holes with higher derivative corrections, in which case the Ryu-Takayanagi prescription no longer applies (see e.g. \cite{Dong:2013qoa}). Remarkably, a class of these theories admit charged black hole solutions with negative entropy \cite{Anninos:2008sj}; it would be desirable to understand the implications for entanglement entropy. Finally, one could consider collapsing black hole solutions with charge. Numerical explorations in \cite{Caceres:2012em,Caceres:2014pda} found that the presence of a chemical potential has a non-trivial effect on the evolution of entanglement entropy, and in some cases can speed up the thermalization. It would be interesting to understand this phenomenon in a controlled analytical expansion \cite{Kundu:2016cgh}.

\section*{Acknowledgements}
It is a pleasure to thank S.~Lokhande and G.~Oling for discussions and comments on the manuscript.
SK is supported by the NSF grant PHY-1316222. JFP is supported by the Foundation for Fundamental Research on Matter (FOM) which is part of the Netherlands Organization for Scientific Research (NWO). JFP would also like to thank the Departament de F\'isica Fonamental at Universitat de Barcelona for the warm hospitality during the final stages of this work.

\appendix

\section{List of all numerical factors\label{factors}}
\begin{align}
a=\frac{2(d-1)}{d-2}\ , \qquad b=\frac{d}{d-2}\ ,
\end{align}
\begin{align}
\S_0=& \frac{2^{d-2} \pi ^{\frac{d-1}{2}} \Gamma \left(-\frac{d-2}{2 (d-1)}\right) }{(d-1) \Gamma \left(\frac{1}{2 (d-1)}\right)} \left(\frac{\Gamma
\left(\frac{d}{2 (d-1)}\right)}{\Gamma \left(\frac{1}{2 (d-1)}\right)}\right)^{d-2}\ , \label{s0}\\
\S_1=& \frac{\Gamma \left(\frac{1}{2 (d-1)}\right)^{d+1}2^{-d-1} \pi ^{-\frac{d}{2}}}{\Gamma \left(\frac{d}{2(d-1)}\right)^d\Gamma
\left(\frac{1}{2}+\frac{1}{d-1}\right)} \left(\frac{\Gamma \left(\frac{1}{d-1}\right) }{\Gamma \left(-\frac{d-2}{2 (d-1)}\right)}+\frac{2^{\frac{1}{d-1}} (d-2)
\Gamma \left(1+\frac{1}{2 (d-1)}\right) }{\sqrt{\pi } (d+1)}\right)\ .\label{s1}
\end{align}
\begin{align}\label{shigh}
\S_{high}=2\left[\frac{\sqrt{\pi } \Gamma \left(-\frac{d-2}{2 (d-1)}\right)}{2 (d-1) \Gamma \left(\frac{1}{2
(d-1)}\right)}\right]+2\int_{0}^{1}dx\left(\frac{\sqrt{1-x^{2(d-1)}}}{x^{d-1}\sqrt{1-x^d}}-\frac{1}{x^{d-1}\sqrt{1-x^{2(d-1)}}}\right)\ .
\end{align}
\begin{align}
N_0&=2\left[\frac{\sqrt{\pi } \Gamma \left(-\frac{d-2}{2 (d-1)}\right)}{2 (d-1) \Gamma \left(\frac{1}{2
(d-1)}\right)}\right]+2\int_{0}^{1}dx\left(\frac{\sqrt{1-x^{2(d-1)}}}{x^{d-1}\sqrt{1-a x^d+b x^{2(d-1)}}}-\frac{1}{x^{d-1}\sqrt{1-x^{2(d-1)}}}\right)\nonumber \\
N_1&=\int_{0}^{1}dx\left(\frac{x\sqrt{1-x^{2(d-1)}}}{\sqrt{1-a x^d+b x^{2(d-1)}}}\right)\left(\frac{1-x^{d-2}}{1-a x^d+b x^{2(d-1)}}\right)\ .\label{n1}
\end{align}

\section{AdS-RN$_{2+1}$\label{AdSRN3}}
Black hole solutions to (\ref{eom1}) and (\ref{eom2}) for $d\ge 2$ take the following form
\begin{eqnarray} \label{RN2}
&& ds^2 = \frac{L^2}{z^2} \left(- f(z) dt^2 + \frac{dz^2}{f(z)} + dx^2 \right) \ , \nonumber\\
&& f(z) = 1-  M z^2 + \frac{Q^2 z^2}{L^2} \log \left( z/ L\right) \ , \\
&& A_t = Q \log\left(z_H/z\right) \ .
\end{eqnarray}
The solution for $d=2$ has peculiar properties: the fall-off of the fields is slower than the standard case and identification of the source and the VEV are subtle \cite{Jensen:2010em}. In particular, the chemical
potential should be identified instead as the sub-leading term as $z\to 0$:
\be
\mu = \frac{Q}{L} \log \left( z_H / L\right) \ .
\ee
A different proposal based on alternative boundary conditions was recently given in \cite{Perez:2015kea}. For simplicity we do not consider this case in our analyses, until these issues are settled.

\section{Extremal surfaces near the horizon}\label{approach}
In this paper, we have exploited the fact that for $\T\l\gg 1$, $z_*\sim z_H$. In this section, let us show how fast $z_*$ approaches $z_H$ in the limit $\T\l\gg 1$. Behavior of the extremal surfaces near the horizon is different for extremal and non-extremal black holes, so, we will discuss them separately.

\subsection{Non-extremal case}
 Assuming that $z_*=z_H(1-\epsilon)$ with $\epsilon\ll1$, from equation (\ref{ellint}),  one can show that
\be
\l=\frac{2d}{4\pi\T}\left[-\frac{ \log \epsilon}{\sqrt{2} \sqrt{(1-d) (d (\varepsilon-2)-2 \varepsilon+2)}}+t_4(d)\right]\ ,
\ee
where $t_4(d)$ is some numerical constant that we will not transcribe here. Therefore, at the leading order
\be\label{expapp}
\epsilon=C_d \exp \left(-\alpha_d(\varepsilon) \T \l \right)\ ,
\ee
where, $C_d$ is some $\O(1)$ numerical factor and
\be
\alpha_d(\varepsilon)=\frac{2\pi}{d}\sqrt{2} \sqrt{(1-d) (d (\varepsilon-2)-2 \varepsilon+2)}\ .
\ee
Therefore, the extremal surface approaches the horizon exponentially fast but it always stays at a finite distance above the horizon. Note that $\alpha_d(\varepsilon)=0$ for the extremal case. One can also check that for the purely thermal case, equation (\ref{expapp}) agrees with the result of \cite{Fischler:2012ca}.

\subsection{Extremal case}

In this case, $T=0$ and the effective temperature is given by (\ref{k1}). Again, assuming that $z_*=z_H(1-\epsilon)$ with $\epsilon\ll1$, from equation (\ref{ellint}),  one can show
\be
\l=\frac{2d}{4\pi\T}\left[\frac{\sqrt{2}}{(d-1) \sqrt{d} \sqrt{\epsilon }}+\frac{\sqrt{2}}{(1-d) \sqrt{d}}+\tau_4(d)\right]\ ,
\ee
where $\tau_4(d)$ is another numerical constant
\be
\tau_4(d)=\int_0^{1}dx \left[\frac{1}{\sqrt{f(x z_H)[(1/x)^{2(d-1)}-1]}}-\frac{1}{\sqrt{2 d} (d-1) (1-x)^{3/2}}\right]_{\varepsilon=a}\ .
\ee
Therefore, at the leading order
\be
\epsilon=\frac{d}{2 \pi ^2 (d-1)^2 \l^2 \T^2}=\frac{8}{(d-1)(d-2)^2\mu^2 \l^2}\ .
\ee
Therefore, for the extremal case, in the limit $\mu\l\gg1$ the extremal surfaces approach the horizon only at a power law rate.


\end{document}